\documentclass[journal]{IEEEtran}

\usepackage{cite}
\usepackage{graphicx}
\usepackage{dcolumn}
\usepackage{bm}
\usepackage{amssymb}
\usepackage{amssymb}

\def\>{\rangle}

\begin{document}
%
\title{Quantum Coherent Nonlinear Feedback with Applications to Quantum Optics on Chip}
%
%
\author{Jing~Zhang, Re-Bing Wu ~\IEEEmembership{Member, IEEE}, Yu-xi Liu, Chun-Wen~Li, Tzyh-Jong~Tarn ~\IEEEmembership{Life Fellow, IEEE}
\thanks{Manuscript received ; first revised ; second revised ; third revised .
        This research was supported in part by the National Natural
        Science Foundation of China under Grant Nos. 61174084, 61134008, 10975080, 61025022, 60904034,
        60836001. T.~J. Tarn would also like to acknowledge partial support from the
        U.S. Army Research Office under Grant W911NF-04-1-0386.}
\thanks{J. Zhang, R.-B. Wu, and C.-W. Li are with the Department of Automation, Tsinghua University, Beijing
100084, P. R. China.(e-mail: jing-zhang@mail.tsinghua.edu.cn;
rbwu@tsinghua.edu.cn; lcw@mail.tsinghua.edu.cn)}
\thanks{Y.-X. Liu is with the Institute of Microelectronics, Tsinghua University,
Beijing 100084, P. R. China.(e-mail:
yuxiliu@mail.tsinghua.edu.cn)}
\thanks{T.-J. Tarn is with the Department of Electrical and Systems Engineering, Washington
University, St. Louis, MO 63130, USA.(e-mail:
tarn@wuauto.wustl.edu)}
\thanks{All the authors are also with the Center for Quantum
Information Science and Technology, Tsinghua National Laboratory
for Information Science and Technology, Beijing 100084, P. R.
China.}}
%
%
%
\markboth{Submitted to Special Issue of IEEE TRANSACTIONS ON
AUTOMATIC CONTROL}{Shell \MakeLowercase{\textit{et al.}}: Bare
Demo of IEEEtran.cls for Journals}
%



\maketitle

\begin{abstract}
In the control of classical mechanical systems, feedback has been
applied to the generation of desired nonlinear dynamics, e.g., in
chaos control. However, how much this can be done is still an open
problem in quantum mechanical systems. This paper presents a
scheme of enhancing nonlinear quantum effects via the recently
developed coherent feedback techniques, which can be shown to
outperform the measurement-based quantum feedback scheme that can
only generate pseudo-nonlinear quantum effects. Apart from the
advantages of our method, an unsolved problem is that the
decoherence rate is also increased by the quantum amplifier, which
may be solved by introducing, e.g., an integral device or an
nonlinear quantum amplifier. Such a proposal is demonstrated via
two application examples in quantum optics on chip. In the first
example, we show that nonlinear Kerr effect can be generated and
amplified to be comparable with the linear effect in a
transmission line resonator (TLR). In the second example, we show
that by tuning the gains of the quantum amplifiers in a TLR
coherent feedback network, the resulting nonlinear effects can
generate and manipulate non-Gaussian ``light" (microwave field)
which exhibits fully quantum sub-Poisson photoncount statistics
and photon antibunching phenomenon. The scheme opens up broad
applications in engineering nonlinear quantum optics on chip.
Particularly, in this study, the concept of feedback
nonlinearization which is very useful for quantum feedback control
systems is introduced. This is in contrast to the feedback
linearization concept used in classical nonlinear feedback control
systems.
\end{abstract}

\begin{keywords}
Feedback nonlinearization, quantum coherent feedback control,
nonlinear quantum optics, on-chip quantum optics, quantum control.
\end{keywords}

%
\IEEEpeerreviewmaketitle

\section{Introduction}\label{s1}
%
%
%
%
\PARstart{O}ver the last decades, the control of quantum phenomena
has been steadily advanced in many fields such as quantum
communication and computation, laser-induced chemical reaction,
and nano
electronics~\cite{Alessandro,HMWisemanbook,HMabuchi_NKhaneja,Rouchon,Brif,DYDong,Huang,Khaneja,Li,Li2,Yuan,Bloch,Albertini,Romano,Yi,Fu,Schirmer,Mirrahimi,Mirrahimi2,NYamamoto,NYamamoto2,DYDong2,Altafini,Altafini2,Altafini3,Alessandro1,Kashima,Bonnard,Wang,Ticozzi,MZhang,WCui,Zhang3,BQi,Geremia,Handel,Jacobs2,Bouten,JKStockton}.
However, the implementation of realtime feedback, which is the
core of control theory and engineering, is still in its infancy.
The major technical obstacles are (1) the time scale of general
quantum dynamics is too fast to be manipulated in realtime by
currently available electronic devices; (2) the required quantum
measurements are generally extremely hard to do; and (3), more
essentially, the back action brought by the quantum measurement
deeply annoys the control designers because it keeps dumping
entropy into the system before the feedback attempts to reduce it.
So far, it is still not clear to what extent quantum feedback
control may outperform the open-loop control, and this impedes the
discoveries of new applications of quantum feedback control.

Up to now, there have been two commonly studied classes of quantum
feedback strategies in the literature. The first one is called the
measurement-based feedback
control~\cite{Wiseman1,Wiseman2,Mabuchi2,Doherty,Mancini1,Mancini2,Mancini3}.
A typical implementation of such a scheme is to shoot an
electromagnetic probe field through the quantum system to be
controlled, carrying a part of information of the system that can
be detected by some measurement apparatus and converted into
classical signals, which are then fed back to adjust the input of
the system. This is the analog of classical feedback loop.
However, the accompanied problem is that the quantum measurement
disturbs the system (i.e., the measurement back-action), and thus
adds unremovable noises. Such a feedback was shown in our
study~\cite{Zhang} to be only capable of generating ``classical"
nonlinearity, which is not fully quantum.

By contrast, the other strategy is called coherent feedback
control~\cite{Lloyd,Nelson,Mabuchi,James,Gough}, in which the
probe field is, instead of being read out, coherently guided back
into the system after being unitarily transformed via quantum
controllers (e.g., quantum beam splitters, quantum switchers, and
quantum amplifiers). Such a strategy preserves the quantum
coherence of the system, and is completely new to the theory of
control system.

This paper will propose a promising application of quantum
feedback control that is capable of inducing fully quantum
nonlinear effects into the controlled systems. Similar idea has
been applied in the pioneer work~\cite{Jacobs3} for a novel
approach to engineer the nonlinear dynamics of the nanomechanical
system by nonlinear feedback control. This is important because
nonlinear quantum processes~\cite{Bajer} are essential in
engineering many interesting quantum phenomena, such as the photon
blockade induced by nonlinear Kerr effect and the generation of
non-Gaussian light, with broad applications in quantum information
processing, quantum nondemolition measurement, and the preparation
of particular nonclassical states, e.g., the Schr\"{o}dinger cat
state. However, the nonlinear effects induced by the natural
field-matter interactions are normally very weak. Therefore,
artificial enhancement of quantum nonlinearity is crucial. Note
that a similar idea has been employed in classical control of
chaos, but the case for quantum control has been rarely
studied~\cite{Jacobs3}, and, as will be seen below, is much more
complicated due to the so-called quantum coherence.

As a potential application, our coherent feedback nonlinearization
scheme can be applied to the emerging field of quantum optics on
chip, i.e., producing optical-like quantum phenomena on compact
solid-state chips such as waveguide circuits or superconducting
circuits~\cite{Sansoni,Politi,Matthews,Berry}. The on-chip optical
devices are powerful in demonstrating particular optical phenomena
that are hard to be observed in conventional optical setups. So
far, the existing on-chip optical experiments are mainly done in
the linear regime, due to the fact that the natural nonlinear
effects induced by the couplings between the optical fields (the
microwave fields) and the solid-state devices are too weak to be
observed. In this paper, we will show that such nonlinear effects
of on-chip lights can be ``artificially" generated and amplified
via the coherent feedback strategy.

This paper is organized as follows. In Sec.~\ref{s2},
preliminaries are given for a brief introduction of the theory of
the coherent feedback control network. In Sec.~\ref{s3}, we
introduce the basic setup and the mathematical model of the
coherent amplification-feedback loop, from which the Hamiltonian
of the controlled system can be effectively reconstructed. In
Sec.~\ref{s4}, we apply these general results to the generation of
the Kerr effect and the cross Kerr effect in the superconducting
circuits. Furthermore, in Sec.~\ref{s5}, we study how to construct
general fourth-order controllable nonlinear quantum Hamiltonians
and their applications to nonclassical on-chip ``lights"
(microwave fields). Conclusions and perspectives are given in
Sec.~\ref{s6}.

\section{Preliminaries}\label{s2}
In quantum mechanics, the state of an isolated system can be
described by a vector $|\psi\rangle$ in an abstract Hilbert space
$\mathcal{H}$, on which the system observables can be described by
operators. The evolution of the system state is governed by the
Schr\"{o}dinger equation:
$$|\dot{\psi}\left(t\right)\rangle=-iH|\psi\left(t\right)\rangle,$$ where the Hamiltonian $H$ is an Hermitian operator denoting the ``energy"
observable of the quantum system. Here, the Planck constant
$\hbar$ has been assigned to be $1$. If the system is bathed with
the external environment, then, instead of the state vector
$|\psi\rangle$, the system state should be represented by the
so-called density operator $\rho$, which is Hermitian and positive
semi-definite on $\mathcal{H}$. The evolution of $\rho$ can be
described by the following master equation~\cite{Puri}:
$$\dot{\rho}\left(t\right)=-i\left[H,\rho\left(t\right)\right]+\mathcal{L}_D\left[\rho\left(\tau\right)\left|\tau\in\left[0,t\right]\right.\right],$$
where the commutator $\left[\cdot,\cdot\right]$ is defined as
$\left[A,B\right]=AB-BA$. The superoperator
$\mathcal{L}_D\left[\rho\left(\tau\right)\left|\tau\in\left[0,t\right]\right.\right]$
represents the dissipation in the system due to the interaction
with its environment, which depends on the system states in the
whole time interval $\left[0,t\right]$. Under the so-called
Markovian approximation to omit the back action effects from the
environment, we can obtain a time-local dissipation superoperator
$\mathcal{L}_D\left[\rho\left(t\right)\right]$.

Generally, in quantum optics, the quantum electromagnetic field
(e.g., the probe light to be used below) is treated as the
collection of the quantized modes whose Hamiltonian reads
$$H_E=\int_{-\infty}^{\infty}d\omega\,\omega
b_{\omega}^{\dagger}b_{\omega},$$ where $b_{\omega}$ is the
annihilation operator of each quantized field mode with frequency
$\omega$. Each $b_\omega$ acts on the corresponding Hilbert state
space spanned by the Fock states
$|n_{\omega}\rangle,\,n_{\omega}=0,1,2,\cdots$, where
$|n_{\omega}\rangle$ represents the state of the quantized field
mode that contains $n_{\omega}$ photons with frequency $\omega$,
in the way that $b_{\omega}|n_{\omega}\rangle$ is proportional to
$|n_{\omega}-1\rangle$. In addition, $b_{\omega}$ also satisfies
the continuous-variable canonical commutation relationship
$\left[b_{\omega},b_{\omega^{\prime}}^{\dagger}\right]=\delta\left(\omega-\omega^{\prime}\right)$.
Traditionally, we use the time-dependent operator
\begin{eqnarray*}
b(t)=\frac{1}{\sqrt{2\pi}}\int d\omega e^{-i\omega t} b_{\omega}
\end{eqnarray*}
to represent the field consisting of a continuum of field modes,
which satisfies
$\left[b(t),b^\dag\left(t^{\prime}\right)\right]=\delta\left(t-t^{\prime}\right)$.

Next, we give a brief review of the theory of the quantum feedback
network developed
recently~\cite{James,Gough,James2,ZhangGF,Gough3,Nurdin,Nurdin2,Maalouf,Yanagisawa1,Yanagisawa2,Gough2,Gough5,Kerckhoff}.
Consider the general input-output system given in Fig.~\ref{Fig of
the input-output system}. The input of the quantum plant (with
internal Hamiltonian $H$) ${\bf b}_{\rm
in}(t)=[b_1(t),\cdots,b_n(t)]^T$ contains $n$ mutually different
fields $b_i\left(t\right),\,i=1,\cdots,n$, all initially in vacuum
states. The input field ${\bf b}_{\rm in}\left(t\right)$ transmits
through a quantum beam splitter described by an $n\times n$
unitary scattering matrix ${\bf S}$, and then interacts with the
plant through the dissipation channels represented by the
dissipation operators ${\bf L}=\left[L_1,\cdots,L_n\right]^T$. In
the following discussions, we will concentrate on the single input
case, i.e., $n=1$, to simplify our discussions.
\begin{figure}
\centerline{\includegraphics[width=5 cm]{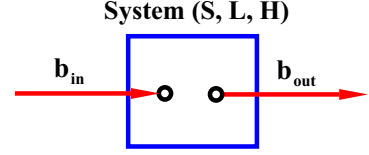}}
\caption{(Color online). The schematic diagram of the input-output
system.}\label{Fig of the input-output system}
\end{figure}

Let $V(t)$ be the unitary evolution operator of the composite
system of the plant plus the input field (as a quantum system).
Correspondingly, the control system can be described by the
following quantum stochastic differential
equation~\cite{Gough2,Hudson}:
\begin{eqnarray}\label{Quantum stochastic differential equation for the system and quantum beam splitter}
\frac{dV(t)}{dt}&=& b_{\rm in}^{\dagger}(t)(S-I)V(t)b_{\rm
in}(t)+b_{\rm in}^{\dagger}(t)L V(t)\nonumber\\
&&-L^{\dagger}S V(t)b_{\rm
in}(t)-\left[\frac{1}{2}L^{\dagger}L-iH\right]V(t),
\end{eqnarray}
with initial value $V_0=I$, where $I$ is the identity operator.
The output field is~\cite{Gardiner}:
\begin{equation}\label{Input-output equality}
b_{\rm out}(t)=V(t)^{\dagger}S V(t)b_{\rm
in}(t)+V(t)^{\dagger}LV(t).
\end{equation}
Equation (\ref{Quantum stochastic differential equation for the
system and quantum beam splitter}) is the Wick-ordered
differential equation of $V_t$ (creators appear on the left,
annihilators on the right), which is equivalent to its quantum
stochastic differential form (see, e.g., Eq.~(30) in
Ref.~\cite{Gough}). Here, $L$ and $H$ ($H$ is self-adjoint) are
system operators commuting with fields $b_{\rm in}\left(s\right)$
and $b^{\dagger}_{\rm in}\left(s\right)$ for earlier times $s<t$,
which means that they are adaptive. $S$ is unitary satisfying
$S^{\dagger}S=SS^{\dagger}=I$, which, for the one-dimensional case
we consider ($n=1$), is just a phase factor
$S=\exp\left(i\theta\right)$. In a compact notation, the above
input-output system can be represented by $\left(S,L,H\right)$.

After averaging over the vacuum input field $b_{\rm in}(t)$ which
can be treated as a quantum random process, the system given by
Eq.~(\ref{Quantum stochastic differential equation for the system
and quantum beam splitter}) can be transformed to the master
equation for the density operator of the plant~\cite{Gough}:
\begin{equation}\label{Master equation of SLH}
\dot{\rho}(t)=-i[H,\rho(t)]+\mathcal{D}\left[L\right]\rho(t),
\end{equation}
where the superoperator $\mathcal{D}\left[L\right]\rho$ is defined
as:
$$\mathcal{D}\left[L\right]\rho=L\rho
L^{\dagger}-\frac{1}{2}L^{\dagger}L\rho- \frac{1}{2}\rho
L^{\dagger}L.$$

Let us consider the two cascade systems shown in Fig.~\ref{Fig of
the series product system}. By introducing the Markovian
approximation to omit the time delay for the output of the first
system $\left(S_1,L_1,H_1\right)$ to reach the second system
$\left(S_2,L_2,H_2\right)$, the total system in the series product
can be described as follows~\cite{Gough}:
\begin{equation}\label{Series product system}
\left(S_2S_1,L_2+S_2L_1,H_1+H_2+\frac{i}{2}\left(L_1^{\dagger}S_2^{\dagger}L_2-L_2^{\dagger}S_2L_1\right)\right).
\end{equation}

\begin{figure}
\centerline{\includegraphics[width=7.5 cm]{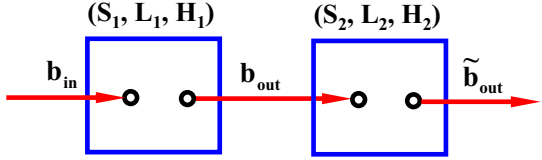}}
\caption{(Color online). The schematic diagram of the series
product system.}\label{Fig of the series product system}
\end{figure}

\begin{figure}
\centerline{\includegraphics[width=5.5 cm]{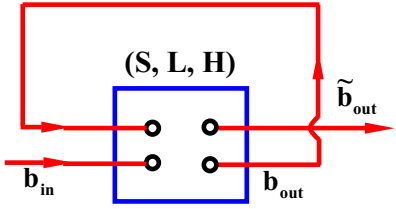}}
\caption{(Color online). The schematic diagram of the coherent
feedback system.}\label{Fig of the coherent feedback system}
\end{figure}

As a special case, if we feed the output of the system
$\left(S,L,H\right)$ back and take it as the input of the same
system to construct a direct coherent feedback network as shown in
Fig.~\ref{Fig of the coherent feedback system}. From
Eq.~(\ref{Series product system}), such a feedback network can be
described by:
\begin{equation}\label{Coherent feedback network}
\left(S^2,L+SL,H+\frac{i}{2}L^{\dagger}\left(S^{\dagger}-S\right)L\right).
\end{equation}

\section{Quantum amplification-feedback loop}\label{s3}

In order to fulfil strong and controllable feedback-induced
nonlinear effects, we consider the modified quantum coherent
amplification-feedback loop shown in Fig.~\ref{Fig for the
coherent feedback and amplification network}. There are two
differences between the traditional coherent feedback network
given in Fig.~\ref{Fig of the coherent feedback system} and the
quantum coherent amplification-feedback loop given in
Fig.~\ref{Fig for the coherent feedback and amplification
network}. Firstly, the dissipation operator $L$ which represents
the interaction between the system and the input field $b_{\rm
in}$ may be different from $L_f$ which represents the interaction
between the system and the feedback field $\tilde{b}_{\rm out}$.
Secondly, we add a quantum amplifier~\cite{Clerk} in the feedback
loop, and feed the output field $b_{\rm out}(t)$ from the system
into it.
\begin{figure}
\centerline{\includegraphics[width=7.5 cm]{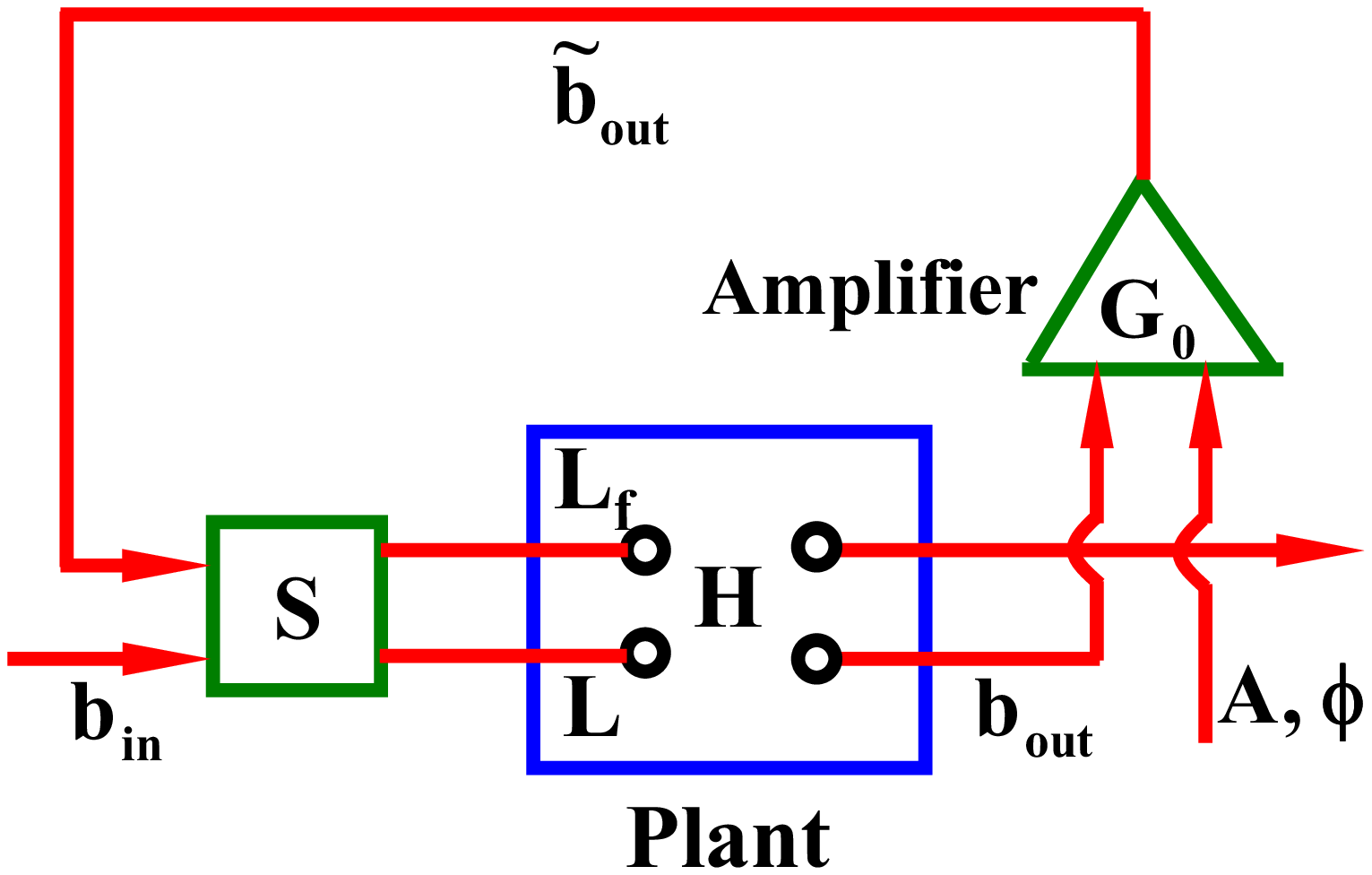}}
\caption{(Color online). The schematic diagram of the coherent
feedback and amplification network.}\label{Fig for the coherent
feedback and amplification network}
\end{figure}

The simplest setup to fulfil such a quantum amplifier is a driven
squeezed cavity field~\cite{Gardiner_Zoller} with the damping rate
$\kappa$ and the tunable squeezing coefficient $\xi$, which can be
realized by the strategy given in Refs.~\cite{Villas-Boas,Almeida}
(see Fig.~\ref{Fig for the quantum amplifier}). As presented in
Refs.~\cite{Villas-Boas,Almeida}, the cavity field is coupled with
a three-level atom which is further driven by a classical field.
By adiabatically eliminating the degrees of freedom of the
auxiliary three-level atom, we can obtain a controllable squeezed
field in the cavity in which the squeezed coefficient is tunable
by changing the coupling strength between the classical driving
field and the three-level atom. This strategy can be extended to
the solid-state superconducting circuit by replacing the cavity by
a transmission line resonator (TLR) and the three-level atom by an
auxiliary flux qubit (see Sec.~V in Ref.~\cite{Zhang2}), which is
used to construct an on-chip amplifier in the following
discussions.

The Hamiltonian of the controllable squeezed cavity field can be
represented, in the rotating frame, as:
\begin{equation}\label{Squeezed cavity field}
H_c=\frac{i\xi}{4}\left(c^{\dagger\,2}-c^2\right)+\sqrt{\kappa}A\left(e^{i\phi}c+c^{\dagger}e^{-i\phi}\right),
\end{equation}
where $c$ is the annihilation operator of the cavity field;
$A,\,\phi\in\mathcal{R}$ represent the normalized amplitude and
the initial phase of the classical control field driving the
cavity mode. In the $\left(S,L,H\right)$ notation, the squeezed
cavity system can be represented by:
\begin{equation}\label{SLH of the squeezed cavity field}
\left(I,\sqrt{\kappa}c,\frac{i\xi}{4}\left(c^{\dagger\,2}-c^2\right)+\sqrt{\kappa}A\left(e^{i\phi}c+c^{\dagger}e^{-i\phi}\right)\right).
\end{equation}
The original system $\left(S,L,H\right)$ and the squeezed field
can be looked as a series product system, which can be represented
from Eq.~(\ref{Series product system}) as:
\begin{equation}\label{SLH of the system and the squeezed field}
\left(S,\sqrt{\kappa}c+L,H+H_c+\frac{i}{2}\sqrt{\kappa}\left(L^{\dagger}c-c^{\dagger}L\right)\right).
\end{equation}
\begin{figure}
\centerline{\includegraphics[width=6 cm]{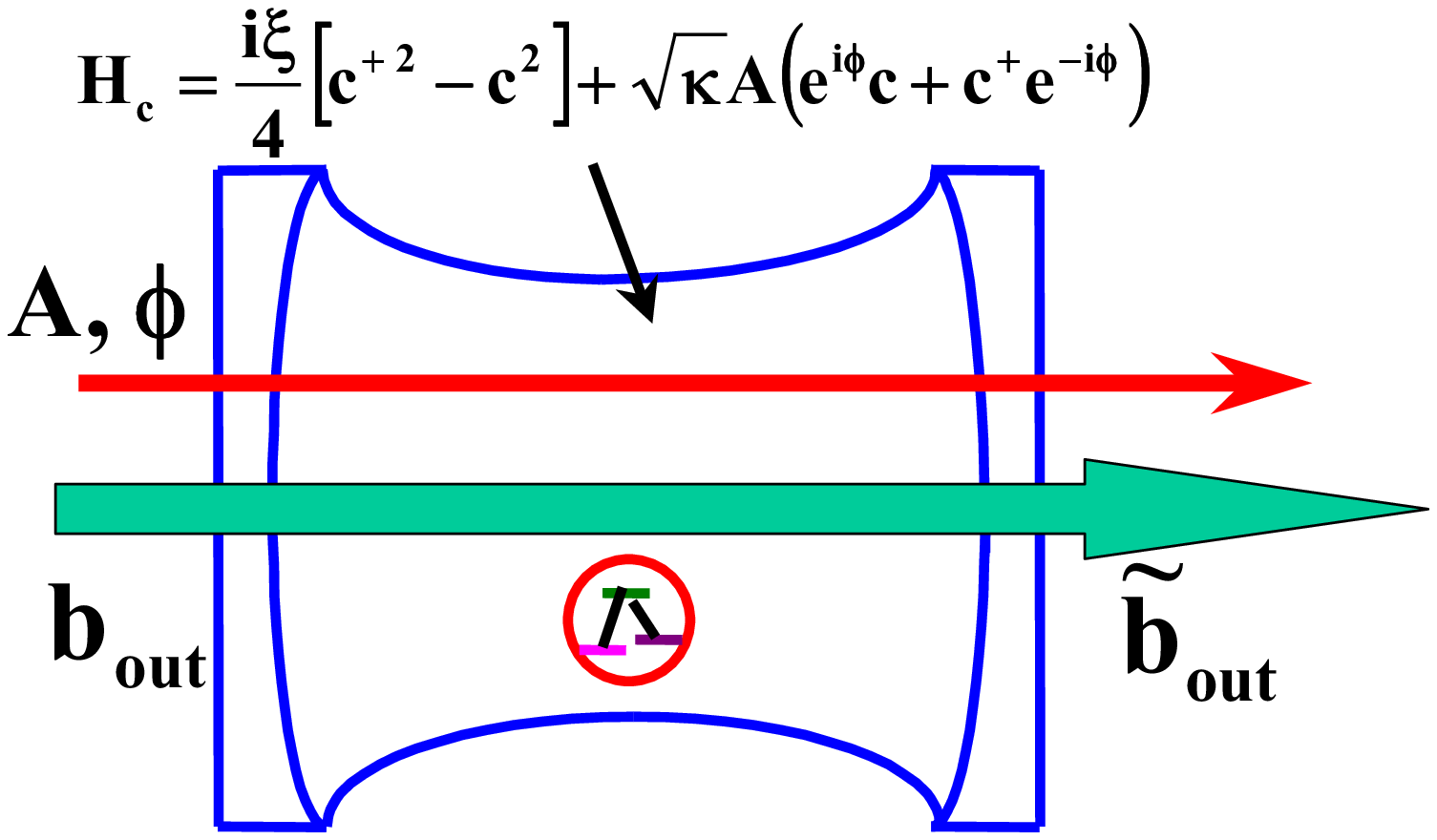}}
\caption{(Color online). The single-mode squeezed field which
works as the quantum amplifier.}\label{Fig for the quantum
amplifier}
\end{figure}
The total coherent amplification-feedback loop given in
Fig.~\ref{Fig for the coherent feedback and amplification network}
can be looked as a series product system of the subsystem
$\left(S,L,H\right)$, the quantum amplifier, and the subsystem
$\left(S,L_f,H\right)$, and thus can be described by:
\begin{eqnarray}\label{SLH of the coherent feedback and amplification network}
&\left(S^2,L_f+S\left(\sqrt{\kappa}c+L\right),H+H_c+\frac{i}{2}\sqrt{\kappa}\left(L^{\dagger}c-c^{\dagger}L\right)\right.&\nonumber\\
&\left.+\frac{i}{2}\left[\left(L^{\dagger}+\sqrt{\kappa}c^{\dagger}\right)S^{\dagger}L_f-L_f^{\dagger}S\left(L+\sqrt{\kappa}c\right)\right]\right).&
\end{eqnarray}

Following the ideas of Refs.~\cite{Gough6,Bouten2}, we can
adiabatically eliminate the degrees of freedom of the cavity mode
by the singular perturbation approach to obtain the following
master equation (see the derivations in Appendix):
\begin{eqnarray}\label{Master equation of the amplification-feedback network}
d\rho&=&-i\left[H+\cosh\left(r_0\right)\left(\frac{i}{2}L_f^{\dagger}SL-\frac{i}{2}L^{\dagger}S^{\dagger}L_f\right)\right.\nonumber\\
&&+\sinh\left(r_0\right)\left(-\frac{i}{4}\left(L^{\dagger}-L^{\dagger}_fS\right)\left(L^{\dagger}+L^{\dagger}_fS\right)+{\rm
h.c.}\right)\nonumber\\
&&+\left\{-\frac{i}{2}Ae^{i\phi}\left[\left(\cosh\left(r_0\right)+1\right)\left(L+S^{\dagger}L_f\right)\right.\right.\nonumber\\
&&\left.\left.\left.+\sinh\left(r_0\right)\left(L^{\dagger}+L^{\dagger}_fS\right)\right]+{\rm
h.c.}\right\},\rho\right]\nonumber\\
&&+\mathcal{D}_s\left[L-S^{\dagger}L_f\right]\rho+\mathcal{D}\left[L-S^{\dagger}L_f\right]\rho,
\end{eqnarray}
where
\begin{eqnarray*}
\mathcal{D}_s\left[\tilde{L}\right]\rho&=&\left(N+1\right)\mathcal{D}\left[\tilde{L}\right]\rho+N\mathcal{D}\left[\tilde{L}^{\dagger}\right]\rho\\
&&+M^*\left(\tilde{L}\rho \tilde{L}-\tilde{L}^2\rho/2-\rho \tilde{L}^2/2\right)\\
&&+M\left(\tilde{L}^{\dagger}\rho
\tilde{L}^{\dagger}-\tilde{L}^{\dagger\,2}\rho/2-\rho
\tilde{L}^{\dagger\,2}/2\right),
\end{eqnarray*}
and
\begin{eqnarray*}
r_0={\rm
ln}\left(\frac{\kappa+\xi}{\kappa-\xi}\right),\,\,N=\frac{\cosh\left(2r_0\right)-1}{2},\,\,M=-\frac{\sinh\left(2r_0\right)}{2}.
\end{eqnarray*}
The two dissipation channels
$\mathcal{D}\left[L-S^{\dagger}L_f\right]\rho$ and
$\mathcal{D}_s\left[L-S^{\dagger}L_f\right]\rho$ are induced by
the vacuum field $b_{\rm in}$ and the squeezed vacuum field
\begin{eqnarray*}
b_{\rm in}^s=\cosh\left(r_0\right)Sb_{\rm
in}-\sinh\left(r_0\right)b_{\rm in}^{\dagger}S^{\dagger},
\end{eqnarray*}
where $b_{\rm in}^s$ is generated by the quantum amplifier from
$b_{\rm in}$, which satisfies that
\begin{eqnarray*}
b_{\rm in}^s\left(t\right)b_{\rm
in}^{s\,\dagger}\left(t^\prime\right)&=&\left(N+1\right)\delta\left(t-t^\prime\right),\\
b_{\rm in}^{s\,\dagger}\left(t\right)b_{\rm
in}^s\left(t^\prime\right)&=&N\delta\left(t-t^\prime\right),\\
b_{\rm in}^s\left(t\right)b_{\rm
in}^s\left(t^\prime\right)&=&M^*\delta\left(t-t^\prime\right),\\
b_{\rm in}^{s\,\dagger}\left(t\right)b_{\rm
in}^{s\,\dagger}\left(t^\prime\right)&=&M\delta\left(t-t^\prime\right),
\end{eqnarray*}
Under the adiabatic approximation, the input-out relation of the
squeezed component can be written as:
\begin{eqnarray*}
\tilde{b}_{\rm out}=\sqrt{G_0}b_{\rm out}+\sqrt{G_0-1}b_{\rm
out}^{\dagger}.
\end{eqnarray*}
It can be seen that the squeezed component in this case can be
looked as a phase-insensitive quantum amplifier
~\cite{Clerk,Gardiner_Zoller} with power gain:
\begin{equation}\label{Power gain of the quantum amplifier}
G_0=\cosh^2\left(r_0\right)=\frac{\left(\kappa^2+\xi^2\right)^2}{\left(\kappa-\xi\right)^2\left(\kappa+\xi\right)^2}.
\end{equation}

The system we discuss cannot be expressed by the
$\left(S,L,H\right)$ notation given in Ref.~\cite{Gough} due to
the existence of the squeezed bath term
$\mathcal{D}_s\left[L-S^{\dagger}L_f\right]\rho$. However, it can
be checked that Eq.~(\ref{Master equation of the
amplification-feedback network}) coincides with those equations in
Ref.~\cite{Gough5} for linear quantum systems, in which linear
quantum dynamical network elements including static Bogoliubov
components (such as squeezers~\cite{Schleier-Smith,Leroux}
discussed here) are formulated by the transfer function and
input-output equation.

If the power gain of the quantum amplifier $G_0$ is far greater
than $1$, the master equation (\ref{Master equation of the
amplification-feedback network}) can be simplified as:
\begin{equation}\label{Master equation when G_0 is far greater than 1}
\dot{\rho}=-i\left[H_{\rm
eff},\rho\right]+G_0\mathcal{D}\left[\frac{1}{2}\left(L-L^{\dagger}+L_f^{\dagger}S-S^{\dagger}L_f\right)\right]\rho,
\end{equation}
where
\begin{eqnarray}\label{Effective Hamiltonian when G_0 is far greater than 1}
H_{\rm
eff}&=&H+\sqrt{G_0}\left(\frac{i}{2}L_f^{\dagger}SL-\frac{i}{2}L^{\dagger}S^{\dagger}L_f\right)\nonumber\\
&&+\sqrt{G_0}\left[-\frac{i}{4}\left(L^{\dagger}-L_f^{\dagger}S\right)\left(L^{\dagger}+L_f^{\dagger}S\right)+{\rm
h.c.}\right]\nonumber\\
&&+\sqrt{G_0}A\cos\phi\left(L+L^{\dagger}+S^{\dagger}L_f+L_f^{\dagger}S\right)
\end{eqnarray}
represents the effective Hamiltonian under coherent feedback. One
can immediately see that the system Hamiltonian has been
reconstructed by the feedback loop involving the following tuple
of parameters:
$${\bf C}=\{S,~L,~L_f,~G_0,~A,~\phi\},$$
which can be properly designed to realize desired quantum dynamics
in the closed-loop system. Note that these parameters can also be
chosen to be time-variant to get more flexibility, but such a case
will not be discussed in this paper.

A system is said to be linear if its Hamiltonian $H$ as a
polynomial of the annihilation and creation operators $a$ and
$a^{\dagger}$ is up to the second-order and the dissipation
operator $L$ is a linear combination of $a$ and $a^{\dagger}$,
otherwise it is said to be nonlinear. Obviously, in Eq.~(\ref{SLH
of the coherent feedback and amplification network}), nonlinear
dynamics can be generated by
\begin{eqnarray}\label{Coherent feedback induced nonlinear Hamiltonian}
H_{nl}&=&\sqrt{G_0}\left[-\frac{i}{4}\left(L^{\dagger}-L_f^{\dagger}S\right)\left(L^{\dagger}+L_f^{\dagger}S\right)+{\rm
h.c.}\right]\nonumber\\
&&+\sqrt{G_0}\left(\frac{i}{2}L_f^{\dagger}SL-\frac{i}{2}L^{\dagger}S^{\dagger}L_f\right),
\end{eqnarray}
if $L$ or $L_f$ is a second-order or higher-order polynomial of
the annihilation and creation operators. What we have discussed
above is the concept of {\it feedback noninearization}. More
importantly, strong nonlinear effects can be produced provided
that the power gain $G_0$ of the quantum amplifier in the coherent
feedback loop is sufficiently high.

In practical experiments, the power gain of the quantum amplifier
cannot be too large. For example, in optical systems, the quantum
amplifiers are typically implemented using a nonlinear optical
material. Thus, the production of a high gain amplifier may
require an optical material with strong nonlinearity which is hard
to be realized. One possible solution of this problem is to
cascade a series of low gain quantum amplifiers to obtain a high
gain just like what we have done for classical systems. There is
also a great progress for nonlinear amplification in solid-state
quantum systems such as superconducting circuits. For example, as
shown in Ref.~\cite{Siddiqi}, the authors found that quantum
signals can be greatly amplified near the bifurcation point of a
nonlinear quantum device, which has been improved by the
succeeding experiments. Thus, it is possible to realize a quantum
amplifier with high gain in solid state systems other than optical
systems.

The idea of the above feedback nonlinearization strategy can be
demonstrated in the following simple model. Let the plant be a
single-mode field with normalized position and momentum operators
as $x=\left(a+a^{\dagger}\right)/\sqrt{2}$ and
$p=\left(-ia+ia^{\dagger}\right)/\sqrt{2}$. The internal
Hamiltonian of the field is $H=\omega a^{\dagger}a$. We choose
$L=\sqrt{\gamma}x^2$, $L_f=\sqrt{\gamma}x$ and $S=e^{i\pi/2}$
respectively. If we coherently feed back the quantum signal (i.e.,
being an operator):
\begin{eqnarray*}
b_{\rm out}(t)=\sqrt{\gamma}x^2+b_{\rm in}(t),
\end{eqnarray*}
where $b_{\rm in}$ is the input vacuum noise, then $H_{nl}$ given
in Eq.~(\ref{Coherent feedback induced nonlinear Hamiltonian})
contains third-order nonlinear terms of $x$ and $p$.

In comparison, the feedback signal in the measurement-based
quantum feedback scheme~\cite{Zhang} is a classical signal (i.e.,
being a scalar):
\begin{eqnarray*}
\bar{b}_{\rm out}(t)=\sqrt{\gamma}\langle x^2\rangle+\xi(t),
\end{eqnarray*}
where $\langle A\rangle={\rm tr}A\rho$ is the average over the
system state $\rho$, and $\xi(t)$ is a classical white noise
satisfying:
\begin{eqnarray*}
E\left(\xi\left(t\right)\right)=0,\quad
E\left(\xi\left(t\right)\xi\left(t'\right)\right)=\delta\left(t-t'\right).
\end{eqnarray*}
It has been shown in Ref.~\cite{Zhang} that such a feedback loop
can only introduce ``classical" nonlinearity for the controlled
quantum system. In fact, statistically, only the trajectories of
the expectation values of the operators $x$ and $p$ behave
nonlinearly. Those of the higher-order quadratures remain
linearly, which can only be altered by fully quantum nonlinear
Hamiltonian terms such as $H_{nl}$ introduced by the coherent
feedback. Such fully quantum nonlinear dynamics is essential to
important physical applications such as light squeezing and the
generation of the Kerr effect.

\section{Generation of Strong Kerr Effects}\label{s4}
In this section, we will introduce the coherent feedback to
generate strong and controllable nonlinear effects in the
following two systems:
\begin{enumerate}
    \item [(1)] Kerr effect in a single-mode field with the annihilation operator $a$, where the nonlinear Hamiltonian
    to be constructed is $H_{\rm Kerr}=\chi\left(a^{\dagger}a\right)^2,\,\chi\in\mathcal{R}$;
    \item [(2)] Cross Kerr effect in a two-mode field with the annihilation
    operators $a$ and $b$, where the nonlinear Hamiltonian is $H_{\rm cross-Kerr}=\chi_{ab}\left(a^{\dagger}a\right)\left(b^{\dagger}b\right),\,\chi_{ab}\in\mathcal{R}$.
\end{enumerate}

The generation of strong Kerr and cross-Kerr effects is crucial to
nonlinear quantum optical phenomena, and has important
applications to the generation of particular quantum states, e.g.,
the Schr\"{o}dinger cat state~\cite{Dodonov} and universal quantum
computation~\cite{Chuang}, i.e., the construction of two-qubit
CNOT gate. Let us see how the Kerr effect can be generated in
on-chip quantum optics~\cite{Sansoni,Politi,Matthews,Berry}
realized by the superconducing circuit~\cite{You,Makhlin,Clarke}
shown in Fig.~\ref{Fig for Kerr and cross-Kerr effects}(a). The
vacuum input field $b_{\rm in}$ transmits through a $\pi/2$ phase
shifter which can be implemented by the on-chip quantum beam
splitter~\cite{Chirolli} proposed recently. Then, the output field
of the phase shifter is coupled to the fundamental mode of the
electric field in a TLR via a charge qubit.

\begin{figure}
\includegraphics[width=5.2 cm]{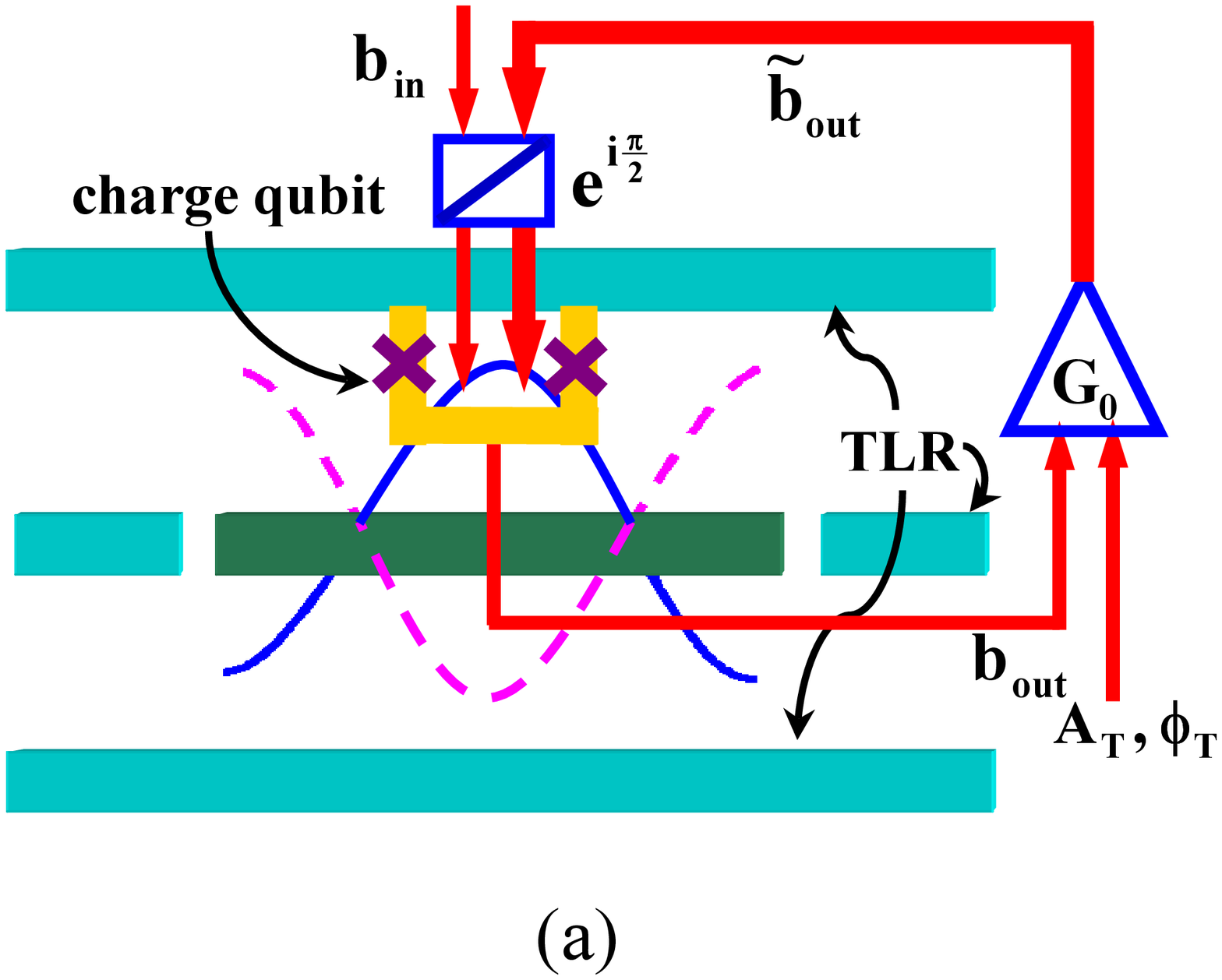}
\includegraphics[width=3.0 cm]{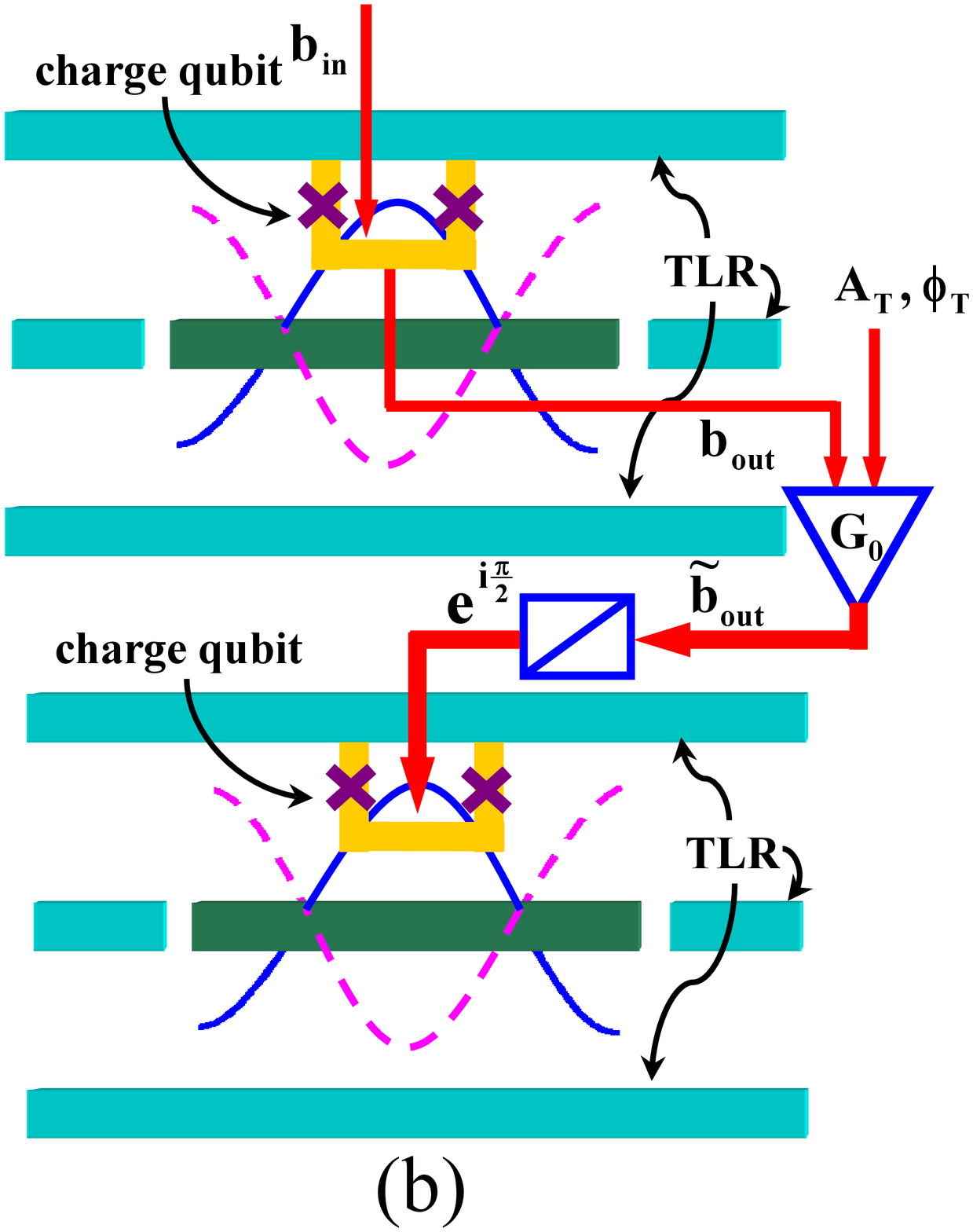}
\caption{(Color online). Schematic diagrams of the superconducting
circuits to generate strong and controllable Kerr and cross-Kerr
effects by coherent feedback and feedforward with (a) for Kerr
nonlinearity and (b) for cross-Kerr nonlinearity.}\label{Fig for
Kerr and cross-Kerr effects}
\end{figure}

The Hamiltonian of the coupled qubit-TLR system can be expressed
as~\cite{You,Makhlin,Clarke}:
\begin{equation}\label{Hamiltonian of the coupled qubit-TLR system}
H_{qT}=4E_C\left(n-n_g\right)^2-2E_J^0\cos\left(\pi\frac{\Phi_x}{\Phi_0}\right)\cos\phi+\omega_a
a^{\dagger}a,
\end{equation}
where $\omega_a$ and $a$ are the angular frequency and the
annihilation operator of the electric field in the TLR; $\phi$ is
a phase operator denoting the phase drop across the
superconducting loop of the charge qubit;
$n=-i\partial/\partial\phi$ is the conjugate operator of $\phi$,
which represents the number of Cooper pairs on the island
electrode of the charge qubit; the reduced charge number $n_g$ on
the gate of the charge qubit, in units of the Cooper pairs, can be
given by $n_g=-C_g V_g/2e$; $C_g$ and $V_g$ are the gate
capacitance and gate voltage; $E_C=e^2/2\left(C_g+2C_J^0\right)$
is the single-electron charging energy of the charge qubit;
$E_J^0$ and $C_J^0$ represent the Josephson energy and the
capacitance of a single Josephson junction; and $\Phi_0$ is the
quantum flux. $\Phi_x$ in Eq.~(\ref{Hamiltonian of the coupled
qubit-TLR system}) denotes the external flux piercing the SQUID
loop of the charge qubit which can be expressed as:
\begin{equation}\label{External flux piercing the SQUID loop of the charge qubit}
\Phi_x=\Phi_e+\eta_T\left(a+a^{\dagger}\right)+\eta_{\rm
in}\left(b_{\rm in}+b_{\rm in}^{\dagger}\right),
\end{equation}
where $\Phi_e$ is the flux generated by the classical magnetic
field through the SQUID loop, and $\eta_T,\,\eta_{\rm in}$ have
units of magnetic flux and their absolute values represent the
strengths of the quantum flux in the SQUID loop induced by the
electric field in the TLR and the input field $b_{\rm in}$. Let
$n_g=1/2$, $\Phi_e=-\Phi_0/2$, and adiabatically eliminate the
degrees of freedom of the charge qubit, we can obtain the
following effective Hamiltonian to represent the coupling between
the TLR and the input field (see the derivations in Appendix):
\begin{equation}\label{Effective Hamiltonian of the coupled TLR and input field system}
H_T=\omega_a a^{\dagger}a-\frac{\pi^3\eta_T^2\eta_{\rm
in}}{\Phi_0^3} a^{\dagger}a\left(b_{\rm in}+b_{\rm
in}^{\dagger}\right).
\end{equation}
Here, we have omitted the linear term of $a$ and $a^{\dagger}$,
which can be compensated by a classical driving field imposed on
the TLR.

With the $\left(S,L,H\right)$ notations, such a system can be
expressed as:
$$H=\omega_a a^{\dagger}a,~ L=\sqrt{\gamma_a}a^{\dagger}a,~
S=e^{i\pi/2},$$
where
\begin{eqnarray*}
\gamma_a=\frac{\pi^6\eta_T^4\eta_{\rm in}^2}{\Phi_0^6}
\end{eqnarray*}
is the damping rate of the TLR induced by the input field. The
output field of the system is fed into a quantum amplifier, and
then fed back into the plant via the damping channel represented
by $L_f=\sqrt{\gamma_a}a^{\dagger}a$.

As analyzed in Sec.~\ref{s3}, the quantum amplifier used here can
be implemented by a squeezed cavity field with a tunable squeezed
coefficient $\xi$, which can be realized by the superconducting
circuit given in Fig.~\ref{Fig for the quantum amplifier in
superconducting circuit}.
\begin{figure}
\centerline{\includegraphics[width=6 cm]{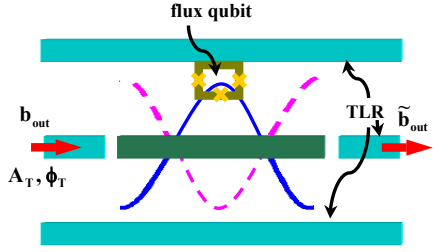}}
\caption{(Color online). Schematic diagram of the quantum
amplifier in superconducting circuit.}\label{Fig for the quantum
amplifier in superconducting circuit}
\end{figure}
In this circuit, a transmission line resonator is driven by the
quantum input field $b_{\rm out}$ and a classical driving field
\begin{eqnarray*}
I\left(t\right)=A_T\cos\left(\omega_T t+\phi_T\right),
\end{eqnarray*}
where the angular frequency of the driving field is equal to the
angular frequency $\omega_T$ of the fundamental mode of the
electric field in the quantum amplifier, i.e., the TLR. The output
field of the quantum amplifier is $\tilde{b}_{\rm out}$. The TLR
is coupled with a flux qubit which works as a $\Delta$-shaped
three-level artificial atom~\cite{Liu}. As discussed in Sec.~V of
Ref.~\cite{Zhang2}, by adiabatically eliminating the degrees of
freedom of the flux qubit, the effective Hamiltonian of the
quantum amplifier can be written under the rotating wave
approximation as:
\begin{eqnarray*}
H_{\rm eff}&=&\omega_T
c^{\dagger}c+\frac{\xi}{4}\left[e^{i\left(\Omega
t+\psi\right)}c^2+c^{\dagger\,2}e^{-i\left(\Omega
t+\psi\right)}\right]\\
&&+A_T\left[e^{i\left(\omega_T
t+\phi_T\right)}c+c^{\dagger}e^{-i\left(\omega_T
t+\phi_T\right)}\right],
\end{eqnarray*}
where $c$ is the annihilation operator of the quantum amplifier,
and $\xi,\,\Omega$, and $\psi$ are tunable parameters. Let
$\Omega=2\omega_T$, $\psi=-\pi/2$, the effective Hamiltonian
$H_{\rm eff}$ can be written in the interaction picture as $H_c$
given in Eq.~(\ref{Squeezed cavity field}). With the
experimentally realizable parameters, the squeezed coefficient
$\xi$ can be as large as the damping rate of the quantum amplifier
$\kappa$ (see, e.g., Ref.~\cite{Zhang2}). Thus, the quantum
amplifier with large power gain $G_0$ given in Eq.~(\ref{Power
gain of the quantum amplifier}) can be obtained if we tune the
squeezed coefficient $\xi$ such that $\xi\approx\kappa$.

With this setup and under the condition that $G_0\gg 1$, an
nonlinear Hamiltonian of the closed-loop dynamics of the
superconducting circuit shown in Fig.~\ref{Fig for Kerr and
cross-Kerr effects}(a) can be reconstructed:
\begin{equation}\label{SLH for the Kerr nonlinearity case}
\tilde{H}=\left(\omega_a-\delta\right)a^{\dagger}a+\chi\left(a^{\dagger}a\right)^2,
\end{equation} where $\delta=2A_T\sqrt{G_0\gamma_a}$ and
$\chi=2\sqrt{G_0}\gamma_a$ are the angular frequency shift and the
strength of the nonlinear Kerr effect induced by the coherent
feedback control respectively. As shown in Eq.~(\ref{SLH for the
Kerr nonlinearity case}), the Kerr effect is enhanced by
increasing the power gain $G_0$, which can be done with the
on-chip amplification device~\cite{Clerk}. As a numerical example,
if the parameters $\omega_a/2\pi=500$ MHz, $\gamma_a/2\pi=1$ MHz,
$A_T^2/2\pi=576$ MHz, and the power gain of the quantum amplifier
$G_0=100$, then it can be calculated that
$$\left(\omega_a-\delta\right)/2\pi=\chi/2\pi=20{\rm MHz}.$$ The strength of the generated Kerr term, which is comparable with that of the
lower-order term, is about $10^4-10^5$ stronger than the Kerr
effect induced by the natural coupling between the electric field
in TLR and the nonlinear element in superconducting circuit (only
around tens of kHz). Thus, with experimentally realizable
parameters, the coherent feedback strategy may dramatically
enlarge the nonlinear Kerr effect.

Furthermore, using the same idea, if the plant includes another
superconducting circuit whose TLR annihilation operator is $b$
(see Fig.~\ref{Fig for Kerr and cross-Kerr effects}(b)), we can
feed the amplified output field into this circuit via
$L_f=\sqrt{\gamma_b}b^{\dagger}b$, and a cross-Kerr Hamiltonian is
obtained as below:
$$H_{\rm cross-Kerr}=\chi_{ab}a^\dag a b^\dag b,$$
whose strength $\chi_{ab}=2\sqrt{G_0\gamma_a\gamma_b}$ can also be
enhanced by increasing the amplification gain $G_0$.

\section{Controllable Fourth-order Nonlinear Dynamics}\label{s5}

This section will focus on the design of a controllable
fourth-order TLR Hamiltonian:
\begin{equation}\label{Controllable four-order Hamiltonian}
H_{\rm eff}=\omega_a a^{\dagger}a+\sum_{k=1}^4\chi_k x_a^k,
\end{equation}
where $x_a=\left(a^{\dagger}+a\right)/\sqrt{2}$ is the normalized
position operator of the TLR; and $\chi_k,\,k=1,2,3,4$ are the
coefficients of the $k$-th order quadratures, which are all
tunable parameters. $H_{\rm eff}$ given in Eq.~(\ref{Controllable
four-order Hamiltonian}) can be used to produce more interesting
nonlinear quantum effects. The terms in Eq.~(\ref{Controllable
four-order Hamiltonian}) have different applications, e.g., the
$\chi_2$-term can be used to realize controllable squeezing in
TLR~\cite{Gough3}; the $\chi_3$-term can be used to construct the
cubic phase gate which is fundamental to realize universal
continuous variable quantum computation~\cite{Lloyd2}; and the
$\chi_4$ term is useful for generating the Kerr effect. More
importantly, as shown below, the nonlinear Hamiltonian given in
Eq.~(\ref{Controllable four-order Hamiltonian}) can be used to
generate non-Gaussian ``light" (microwave field) to show fully
quantum sub-Poisson photoncount statistics and photon antibunching
phenomenon~\cite{Bajer}. The non-Gaussian ``light" generated is
possible to be used to transmit quantum information, which may
have higher capacity of the information transmission than the
Gaussian light in continuous variable quantum
communication~\cite{Braunstein}.

\begin{figure}
\includegraphics[bb=23 323 565 771, width=4.8 cm, clip]{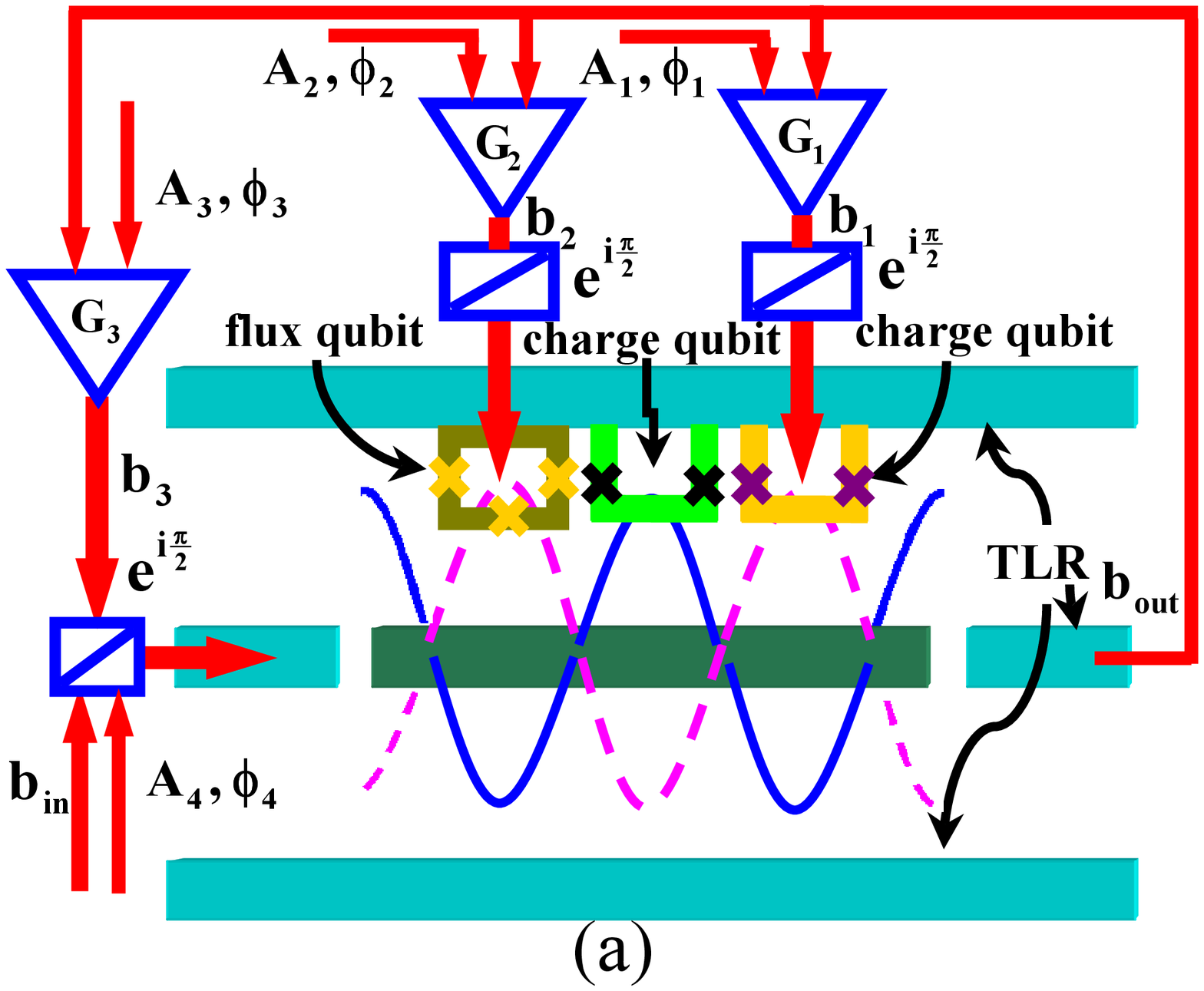}
\includegraphics[bb=119 210 415 529, width=3.5 cm, clip]{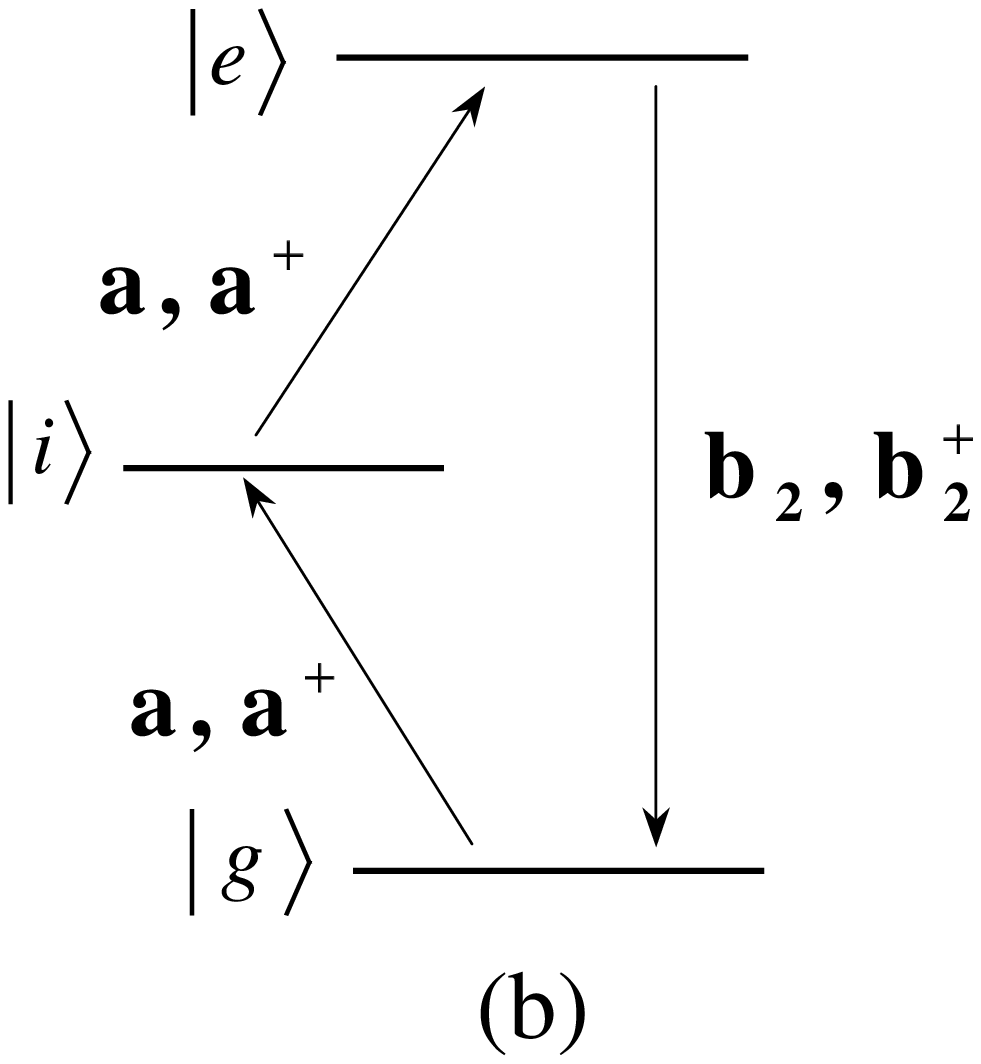}
\caption{(Color online). (a) Schematic diagram used to generate
the fourth-order controllable Hamiltonian given in
Eq.~(\ref{Controllable four-order Hamiltonian}). (b) The
$\Delta$-shape transition of the auxiliary flux qubit.} \label{Fig
for fulfilling controllable four-order Hamiltonian}
\end{figure}

Such a nonlinear Hamiltonian can be constructed via the
superconducting circuit in Fig.~\ref{Fig for fulfilling
controllable four-order Hamiltonian}(a), where a TLR is coupled
with a flux qubit (left) and two charge qubits (central, and
right). The central charge qubit works as an auxiliary device for
the quantum detection. By tuning the parameters of this auxiliary
charge qubit, we can execute a detection of the square of the
normalized position operator $x_a$ by the probe field through the
TLR. With this detection, we can obtain a quadratic damping
operator $L=\sqrt{\gamma}x_a^2$, where $\gamma$ is the related
damping rate. This is similar to the quantum measurement strategy
given in Ref.~\cite{Jacobs} to detect the square of the normalized
position operator of a nanomechanical resonator. The charge qubit
on the right plays the same role as in Sec.~\ref{s4} to induce a
nonlinear Kerr term. The flux qubit in the circuit can be treated
as a three-level artificial atom with $\Delta$-shape
transition~\cite{Liu,Deppe}, whose interaction with the electric
field in the TLR induces another nonlinear Hamiltonian. As shown
in Fig.~\ref{Fig for fulfilling controllable four-order
Hamiltonian}(b), a two-photon exchange process occurs between the
cavity mode $a$ in the TLR and the travelling-wave mode $b_2$ via
the flux qubit, i.e., the $\Delta$-shaped three-level artificial
atom. Two photons in the cavity mode $a$ with the same angular
frequency $\omega_a$ annihilate, and one photon in the
travelling-wave mode $b_2$ is created in this process, and vice
versa. Finally, the output field is coherently fed back after
amplification to drive the electric field in the TLR, which leads
to a third nonlinear Hamiltonian.

Mathematically, the above setup results in three feedback loops
with
\begin{eqnarray*}
 && S_1=S_2=S_3=e^{i\pi/2}, \\
 && L_1=L_2=L_3=\sqrt{\gamma}x_a^2, \\
&&  L_{1f}=\sqrt{\gamma_1}a^{\dagger}a,~
L_{2f}=\sqrt{\gamma_2}a^{\dagger\,2},~L_{3f}=\sqrt{\gamma_3}x_a.
\end{eqnarray*}
The corresponding parameters $A_j,\,\phi_j,\,j=1,2,3,4$ of the
inputs of the quantum amplifiers and the phase shifters are
tunable parameters need to be designed. Here, we let
$\phi_1=\phi_2=0,\,\phi_3=-\pi,\,\phi_4=-\pi/2$. From
Eq.~(\ref{Effective Hamiltonian when G_0 is far greater than 1}),
we can obtain the desired effective Hamiltonian~(\ref{Controllable
four-order Hamiltonian}) with
\begin{eqnarray*}
 &\chi_1=A_4\sqrt{2\gamma},\quad\chi_2=4A_1\sqrt{G_1\gamma_1}-2A_3\sqrt{G_3\gamma_3},& \\
 &\chi_3=2\sqrt{G_3\gamma\gamma_3},\quad\chi_4=2\sqrt{G_1\gamma\gamma_1},&
\end{eqnarray*}
where we have set parameters $G_2=G_1\gamma_1/\gamma_2$ and
$A_2=A_1\sqrt{\gamma_2/\gamma_1}$ in order to obtain the
Hamiltonian form shown in Eq.~(\ref{Controllable four-order
Hamiltonian}). Therefore, by tuning the control parameters
$G_1,\,G_3,\,A_1,\,A_3$, and $A_4$, we can independently change
the coefficients $\chi_k,\,\,k=1,2,3,4$.

These nonlinear terms can be designed to generate nonclassical
microwave field (i.e., the so-called nonclassical ``light") in
TLR. To illustrate the effectiveness of the coherent feedback
scheme, we set the parameters:
\begin{eqnarray*}
  &\omega_a/2\pi=100\,{\rm MHz},~\gamma_1/2\pi=\gamma_3/2\pi=1\,{\rm MHz},&  \\
  &\gamma/2\pi=1\,{\rm MHz},~G_1=G_3=10^3,& \\
  & A_1^2/2\pi=40\,{\rm MHz},~A_3^2/2\pi=152.1\,{\rm MHz}, & \\
  & A_4^2/2\pi=200\,{\rm MHz}. &
\end{eqnarray*}

As shown in Fig.~\ref{Fig for the coherent feedback control
induced on-chip nonlinear optics}(a), we can observe the
sub-Poisson photoncount statistics indicating by the fano factor
\begin{eqnarray*}
F=\left(\langle N_a^2 \rangle-\langle N_a\rangle^2\right)/\langle
N_a \rangle <1,
\end{eqnarray*}
and the photon antibunching phenomenon indicated by
\begin{eqnarray*}
g^{(2)}(\tau)>g^{(2)}\left(0\right),
\end{eqnarray*}
where $N_a=a^{\dagger}a$ is the photon number operator of the TLR;
$\langle\cdot\rangle$ is the average over the system state; and
the normalized second-order correlation function
$g^{\left(2\right)}\left(\tau\right)$ is defined by:
$$g^{(2)}(\tau)=\frac{\langle a^{\dagger}\left(t\right)a^{\dagger}\left(t+\tau\right)a\left(t+\tau\right)a\left(t\right)\rangle}{\langle a^{\dagger}\left(t\right)a\left(t\right)\rangle^2}.$$
Here $a\left(t\right)$ is the operator in the Heisenberg picture
defined by ${\rm tr}\left(a\left(t\right)\rho_0\right)={\rm
tr}\left(a\rho\left(t\right)\right)$, where $\rho_0$ and
$\rho\left(t\right)$ are the initial state of the system and the
state of the system at time $t$. Sub-Poisson photoncount
statistics and photon antibunching phenomenon are typical quantum
phenomena which violate the Cauchy-Schwartz inequality for the
classical lights. Additionally, different from the Gaussian
lights, e.g., the laser, which are quite similar to classical
lights, the on-chip light generated is indeed a non-Gaussian
light, which is highly nonclassical. In Fig.~\ref{Fig for the
coherent feedback control induced on-chip nonlinear optics}(b), we
use the measure
$$\delta\left[\rho\right]=\frac{{\rm
tr}\left[(\rho-\sigma)^2/2\right]}{{\rm
tr}\left[\rho^2\right]}\in[0,1]$$ to evaluate the non-Gaussian
degree of the light generated~\cite{Genoni}, where $\sigma$ is a
Gaussian state with the same first and second-order quadratures of
the non-Gaussian state $\rho$. Simulation results in Fig.~\ref{Fig
for the coherent feedback control induced on-chip nonlinear
optics}(b) show that high-quality non-Gaussian state with
$\delta\left[\rho\right]>0.25$ can be obtained. As pointed out by
Ref.~\cite{Genoni2}, the maximal value of
$\delta\left[\rho\right]$ is not larger than $1/2$ for single-mode
quantum states. Thus, the non-Gaussian degree of the generated
light can be larger than half of the maximal value that can be
reached by any non-Gaussian states.

\begin{figure}
\centerline{\includegraphics[width=4.25 cm]{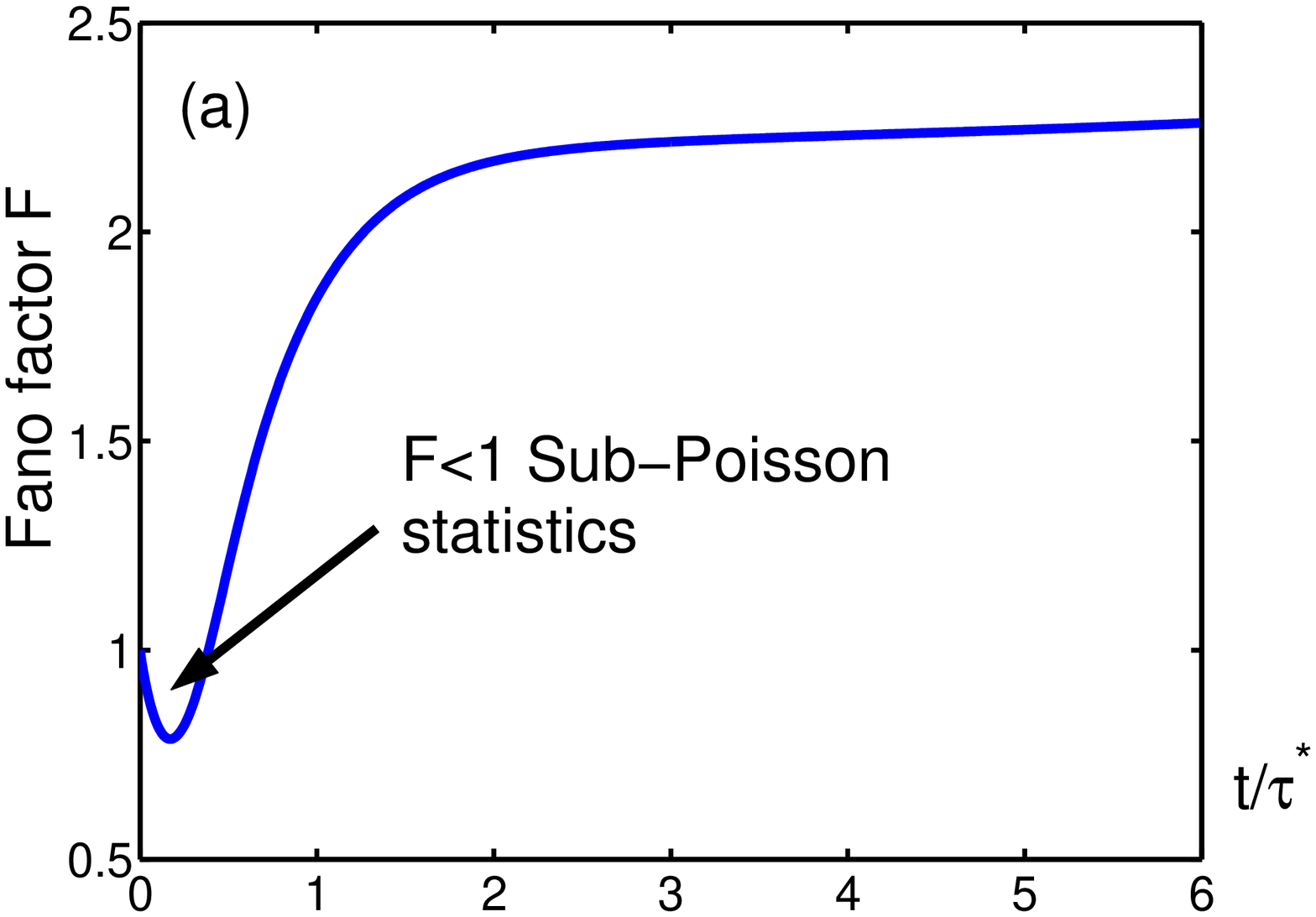}
\includegraphics[width=4.25 cm]{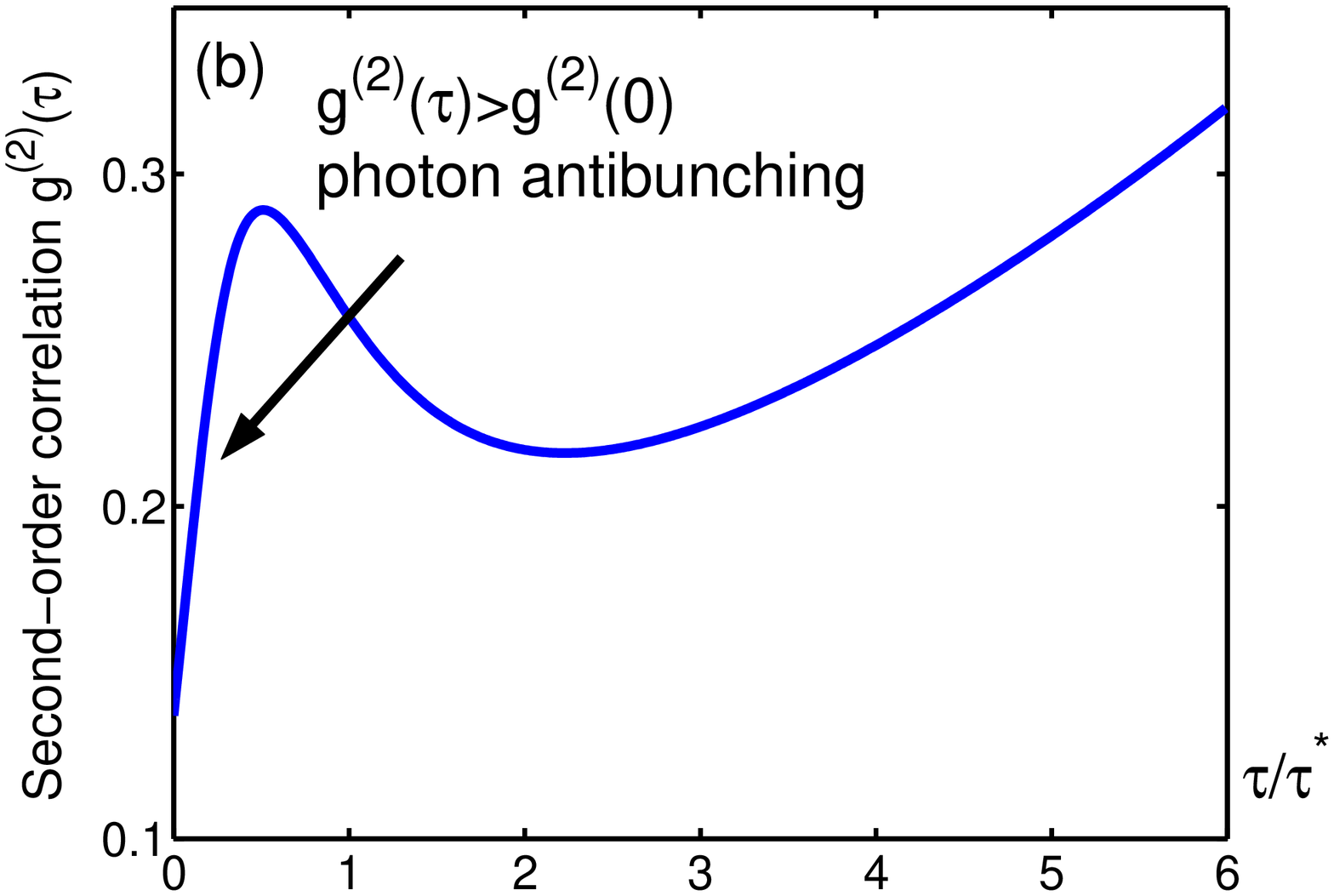}}
\centerline{\includegraphics[width=4.25 cm]{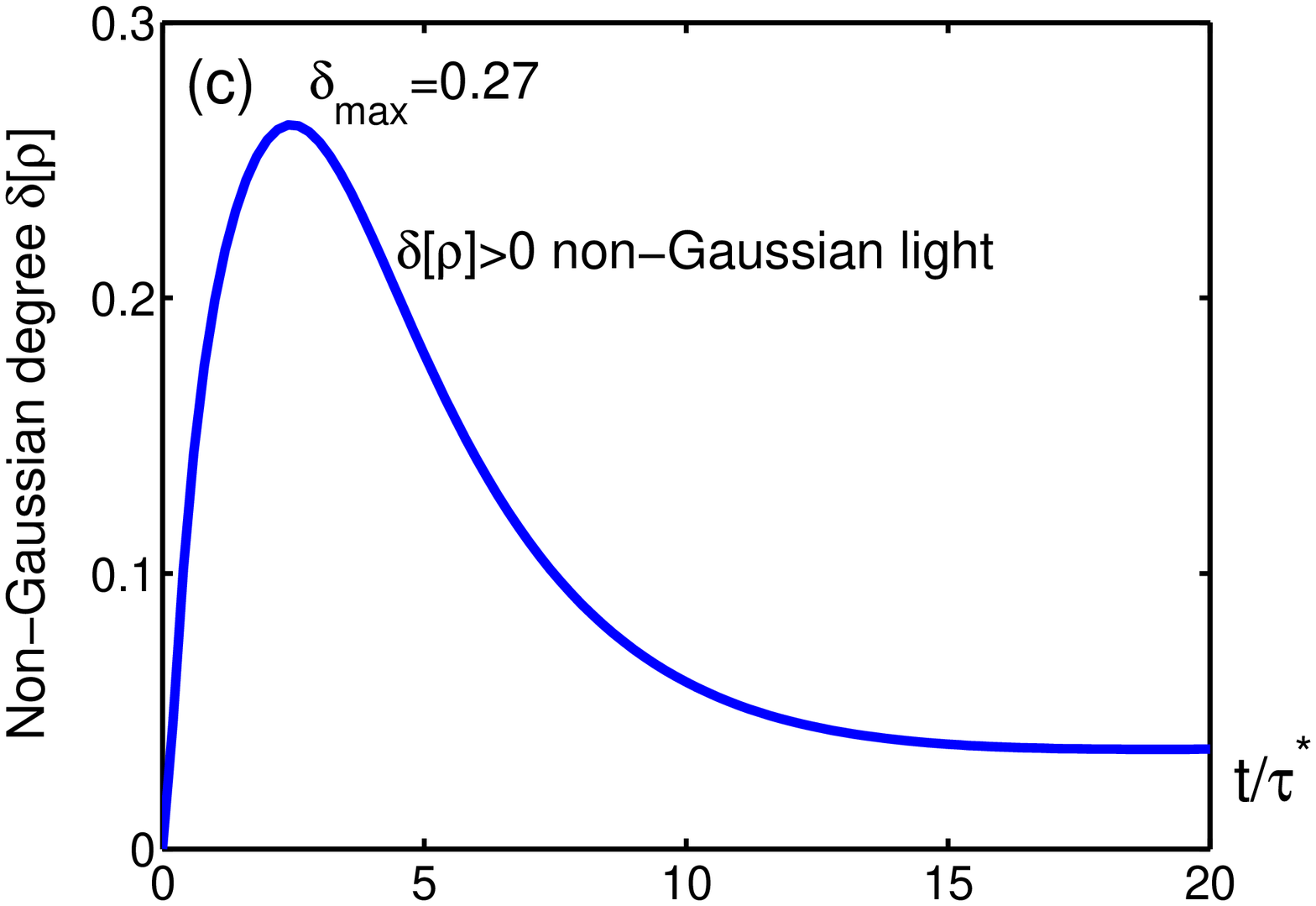}}
\caption{(Color online) (a) Fano factor $F(t)$ and (b) the
normalized second-order correlation function $g^{(2)}(\tau)$: the
sub-Poisson photoncount statistics indicating by $F<1$ and the
photon antibunching indicating by
$g^{(2)}(\tau)>g^{(2)}\left(0\right)$ seen in small time scale
exhibit the nonclassical properties of the light generated. (c)
non-Gaussian measure $\delta\left[\rho\right]$: a highest value
$0.27$ can be attained, and it is shown that the stationary state
of the electric field is also non-Gaussian and thus nonclassical
state. $\tau^*=0.2$ ns is a normalized time-scale.}\label{Fig for
the coherent feedback control induced on-chip nonlinear optics}
\end{figure}

Finally, it should be pointed out that the decoherence (i.e., the
destruction of quantum coherence) rate will also be increased by
the quantum amplifier that is designed to enhance the nonlinear
effect. Thus, there is a tradeoff between pursuing strong
nonlinear effect and weak decoherence, which may somewhat limit
the ability of effectively manipulating nonlinear quantum optical
phenomena, e.g., the generation of Schr\"{o}dinger cat state. This
problem will hopefully be solved by introducing a nonlinear
amplifier whose signal/noise ratio is high.

\section{Conclusion}\label{s6}

In summary, we present a method of engineering strong and
controllable nonlinear effects in quantum systems by coherent
feedback control and amplification. A byproduct of this
investigation is the introduction of the concept of feedback
nonlinearization which is very useful for quantum feedback control
systems. To the authors' knowledge this concept has never been
discussed in the literature. The applications in TLR
superconducting circuits demonstrate its power of generating
strong and controllable nonlinear Kerr and cross-Kerr effects and
more complex nonlinear phenomena. They open up new perspectives to
the design of nonlinear circuits for quantum optics on chip, where
the systematic design methodology of the feedback loop parameters,
including $S$, $G_0$, etc., for more complex nonlinearities and
large-scale circuits are interesting topics to be studied in the
future. There is still a problem left. In our method, the
decoherence will be enhanced if we want to generate stronger
quantum nonlinearity, which may limit our ability to manipulate
nonlinear quantum effects, e.g., the generation of Schr\"{o}dinger
cat state. This problem is also left for the future study.


%
%

\appendix
{\it Derivation of Eq.~(\ref{Master equation of the
amplification-feedback network}):} To adiabatically eliminate the
degrees of freedom of the squeezed cavity field, we introduce the
singular perturbation approach~\cite{Gough6} to let
$\kappa=\kappa_0/\epsilon$, $\xi=\xi_0/\epsilon$, and let
$\epsilon\rightarrow 0$ in the final step. In this case, the total
system composed of the system $\left(S,L,H\right)$, the squeezed
cavity field, and the system $\left(S,L_f,H\right)$ can be
described by:
\begin{eqnarray}\label{SLH of the coherent feedback and amplification network under singular perturbation}
&\left(S^2,L_f+S\left(\sqrt{\frac{\kappa_0}{\epsilon}}c+L\right),H+\frac{i\xi_0}{4\epsilon}\left(c^{\dagger\,2}-c^2\right)\right.&\nonumber\\
&+\sqrt{\frac{\kappa_0}{\epsilon}}A\left(e^{i\phi}c+c^{\dagger}e^{-i\phi}\right)+\frac{i}{2}\sqrt{\frac{\kappa_0}{\epsilon}}\left(L^{\dagger}c-c^{\dagger}L\right)&\nonumber\\
&\left.+\frac{i}{2}\left(L^{\dagger}S^{\dagger}L_f-L_f^{\dagger}SL\right)+\frac{i}{2}\sqrt{\frac{\kappa_0}{\epsilon}}\left(c^{\dagger}S^{\dagger}L_f-L_f^{\dagger}SC\right)\right).&\nonumber\\
\end{eqnarray}
From Eq.~(30) in Ref.~\cite{Gough} and Eq.~(\ref{SLH of the
coherent feedback and amplification network under singular
perturbation}), the dynamical equation of the evolution operator
$U_t\left(\epsilon\right)$ of the total system can be expressed
as:
\begin{eqnarray*}
dU_t\left(\epsilon\right)&=&\left\{\left(S^2-I\right)d\Lambda_t+dB_t^{\dagger}\left(L_f+SL+\sqrt{\frac{\kappa_0}{\epsilon}}Sc\right)\right.\\
&&-\left(L_f^{\dagger}+L^{\dagger}S^{\dagger}+\sqrt{\frac{\kappa_0}{\epsilon}}c^{\dagger}S^{\dagger}\right)S^2dB_t\\
&&-\frac{1}{2}\left(L_f^{\dagger}+L^{\dagger}S^{\dagger}+\sqrt{\frac{\kappa_0}{\epsilon}}c^{\dagger}S^{\dagger}\right)\\
&&\left(L_f+SL+\sqrt{\frac{\kappa_0}{\epsilon}}Sc\right)dt\\
&&-i\left[H+\frac{i}{2}\left(L^{\dagger}S^{\dagger}-L_F^{\dagger}SL\right)+\frac{i\xi_0}{4\epsilon}\left(c^{\dagger\,2}-c^2\right)\right.\\
&&+\sqrt{\frac{\kappa_0}{\epsilon}}\left(\left(\frac{i}{2}\left(L^{\dagger}-L_f^{\dagger}S\right)+Ae^{i\phi}\right)c\right.\\
&&\left.\left.\left.+c^{\dagger}\left(-\frac{i}{2}\left(L-S^{\dagger}L_f\right)+Ae^{-i\phi}\right)\right)\right]dt\right\}U_t\left(\epsilon\right),
\end{eqnarray*}
where
\begin{eqnarray*}
\Lambda_t=\int_0^t b_{\rm in}^{\dagger}\left(\tau\right)b_{\rm
in}\left(\tau\right)d\tau,\,\,\,B_t=\int_0^t b_{\rm
in}\left(\tau\right)d\tau.
\end{eqnarray*}
In order to eliminate the degrees of freedom of the cavity mode in
the singular limit $\epsilon\rightarrow 0^{+}$, we change to the
interaction picture by introducing the time evolution operator
$V_t$, which satisfies that:
\begin{eqnarray}\label{Evolution of V_t}
dV_t\left(\epsilon\right)&=&\left[\left(S^2-I\right)d\Lambda_t+dB_t^{\dagger}\left(L_f+SL+\sqrt{\frac{\kappa_0}{\epsilon}}Sc\right)\right.\nonumber\\
&&-\left(L_f^{\dagger}+L^{\dagger}S^{\dagger}+\sqrt{\frac{\kappa_0}{\epsilon}}c^{\dagger}S^{\dagger}\right)S^2dB_t\nonumber\\
&&-\frac{1}{2}\left(L_f^{\dagger}+L^{\dagger}S^{\dagger}+\sqrt{\frac{\kappa_0}{\epsilon}}c^{\dagger}S^{\dagger}\right)\nonumber\\
&&\left(L_f+SL+\sqrt{\frac{\kappa_0}{\epsilon}}Sc\right)dt\nonumber\\
&&\left.+\frac{\xi_0}{4\epsilon}\left(c^{\dagger\,2}-c^2\right)dt\right]V_t\left(\epsilon\right).
\end{eqnarray}
Then, we want to consider the evolution of the unitary operator
$\tilde{U}_t\left(\epsilon\right)=V_t^{\dagger}\left(\epsilon\right)U_t\left(\epsilon\right)$,
which satisfies that
\begin{equation}\label{Evolution in the interaction picture}
\frac{d\tilde{U}_t\left(\epsilon\right)}{dt}=-i\tilde{H}\left(\epsilon\right)\tilde{U}_t\left(\epsilon\right).
\end{equation}
The Hamiltonian $\tilde{H}\left(\epsilon\right)$ is defined by
$\tilde{H}\left(\epsilon\right)=V_t^{\dagger}\left(\epsilon\right)H\left(\epsilon\right)V_t\left(\epsilon\right)$,
where
\begin{eqnarray*}
H\left(\epsilon\right)&=&\sqrt{\frac{\kappa_0}{\epsilon}}\left\{\left[\frac{i}{2}\left(L^{\dagger}-L_f^{\dagger}S\right)+Ae^{i\phi}\right]c+{\rm
h.c.}\right\}\\
&&+H+\frac{i}{2}\left(L^{\dagger}S^{\dagger}L_f-L_f^{\dagger}SL\right).
\end{eqnarray*}
Then, we introduce the normalized cavity mode in the interaction
picture:
\begin{eqnarray*}
\tilde{b}_t\left(\epsilon\right)=-\sqrt{\frac{\kappa_0}{4\epsilon}}V_t^{\dagger}\left(\epsilon\right)c
V_t\left(\epsilon\right).
\end{eqnarray*}
From Eq.~(\ref{Evolution of V_t}), it can be verified that
\begin{eqnarray}\label{Evolution of btepsilon}
d\tilde{b}_t\left(\epsilon\right)&=&-\frac{\kappa_0}{2
\epsilon}\tilde{b}_t\left(\epsilon\right)dt+\frac{\xi_0}{2\epsilon}\tilde{b}_t^{\dagger}\left(\epsilon\right)dt\nonumber\\
&&+\frac{\kappa_0}{2\epsilon}\left[\frac{1}{2}\left(L+S^{\dagger}L_f\right)dt+SdA_t\right].
\end{eqnarray}
It can be solved from Eq.~(\ref{Evolution of btepsilon}) that
\begin{eqnarray}\label{Solution of btepsilon}
\tilde{b}_t\left(\epsilon\right)&=&\sqrt{\frac{\epsilon}{\kappa_0}}\left[G_{1\epsilon}\left(t\right)\left(c+c^{\dagger}\right)+G_{2\epsilon}\left(t\right)\left(c-c^{\dagger}\right)\right]\nonumber\\
&&+\frac{\kappa_0}{\kappa_0-\xi_0}\int_0^tG_{1\epsilon}\left(t-\tau\right)\left[\frac{1}{2}\left(L+S^{\dagger}L_f\right)d\tau\right.\nonumber\\
&&\left.+SdA_{\tau}+{\rm
h.c.}\right]\nonumber\\
&&+\frac{\kappa_0}{\kappa_0+\xi_0}\int_0^tG_{2\epsilon}\left(t-\tau\right)\left[\frac{1}{2}\left(L+S^{\dagger}L_f\right)d\tau\right.\nonumber\\
&&\left.+SdA_{\tau}-{\rm h.c.}\right],
\end{eqnarray}
where
\begin{eqnarray*}
G_{1\epsilon}\left(t\right)&=&\frac{\kappa_0-\xi_0}{4\epsilon}\exp\left(-\frac{\left(\kappa_0-\xi_0\right)\left|\tau\right|}{2\epsilon}\right),\\
G_{2\epsilon}\left(t\right)&=&\frac{\kappa_0+\xi_0}{4\epsilon}\exp\left(-\frac{\left(\kappa_0+\xi_0\right)\left|\tau\right|}{2\epsilon}\right).
\end{eqnarray*}
From Eq.~(\ref{Evolution in the interaction picture}), we have
\begin{eqnarray}\label{Evolution in the interaction picture with btepsilon}
\frac{d}{dt}\tilde{U}_t\left(\epsilon\right)&=&-i\left\{\left[-i\left(L^{\dagger}-L_f^{\dagger}S\right)-2Ae^{i\phi}\right]\tilde{b}_t\left(\epsilon\right)\right.\nonumber\\
&&+\tilde{b}^{\dagger}_t\left(\epsilon\right)\left[i\left(L-S^{\dagger}L_f\right)-2Ae^{-i\phi}\right]\nonumber\\
&&\left.+H+\frac{i}{2}\left(L^{\dagger}S^{\dagger}L_f-L_f^{\dagger}SL\right)\right\}\tilde{U}_t\left(\epsilon\right)dt.\nonumber\\
\end{eqnarray}
Substituting Eq.~(\ref{Solution of btepsilon}) into
Eq.~(\ref{Evolution in the interaction picture with btepsilon})
and letting $\epsilon\rightarrow 0$ (it is a weak convergence in
the meaning of Ref.~\cite{Gough6}), we have
\begin{eqnarray}\label{Evolution after adiabatically eliminating the cavity mode}
d\tilde{U}_t&=&-i\left\{\left[-\frac{i}{4}\left(L^{\dagger}-L_f^{\dagger}S\right)\left(\cosh\left(r_0\right)\left(L+S^{\dagger}L_f\right)\right.\right.\right.\nonumber\\
&&\left.\left.+\sinh\left(L^{\dagger}+L_f^{\dagger}S\right)\right)+{\rm
h.c.}\right]dt\nonumber\\
&&+\left[-\frac{1}{2}Ae^{i\phi}\left(\left(\cosh\left(r_0\right)+1\right)\left(L+S^{\dagger}L_f\right)\right.\right.\nonumber\\
&&\left.\left.+\sinh\left(r_0\right)\left(L^{\dagger}+L_f^{\dagger}S\right)\right)+{\rm
h.c.}\right]dt+Hdt\nonumber\\
&&-\frac{i}{2}\left(L^{\dagger}-L_f^{\dagger}S\right)dB_t^s+\frac{i}{2}\left(L-S^{\dagger}L_f\right)dB_t^{s\,\dagger}\nonumber\\
&&-\frac{i}{2}\left(L^{\dagger}-L_f^{\dagger}S\right)SdB_t\nonumber\\
&&\left.+\frac{i}{2}\left(L-S^{\dagger}L_f\right)dB_t^{\dagger}S^{\dagger}\right\}\tilde{U}_t,
\end{eqnarray}
where
\begin{eqnarray*}
dB_t^s=\cosh\left(r_0\right)SdB_t+\sinh\left(r_0\right)dB_t^{\dagger}S^{\dagger}.
\end{eqnarray*}
Here, we have used the limit
\begin{eqnarray*}
G_{\epsilon}\left(\tau\right)=\frac{\kappa}{4\epsilon}\exp\left(-\frac{\kappa\left|\tau\right|}{2\epsilon}\right)\rightarrow\delta\left(\tau\right),\,\,\,\epsilon\rightarrow
0.
\end{eqnarray*}
Since $\rho={\rm tr}_E\left[\left(\tilde{U}_t\rho_0
\tilde{U}_t^{\dagger}\right)/{\rm tr}\left(\tilde{U}_t\rho_0
\tilde{U}_t^{\dagger}\right)\right]$ where ${\rm
tr}_E\left(\cdot\right)$ is the partial trace over the Hilbert
space of the noise $dB_t$, we can obtain the master equation
(\ref{Master equation of the amplification-feedback network}) of
$\rho$ from Eq.~(\ref{Evolution after adiabatically eliminating
the cavity mode}) with simple calculations. Note that the map
\begin{eqnarray*}
T\left(t\right):\,\rho_0\mapsto{\rm
tr}_E\left[\left(\tilde{U}_t\rho_0
\tilde{U}_t^{\dagger}\right)/{\rm tr}\left(\tilde{U}_t\rho_0
\tilde{U}_t^{\dagger}\right)\right]
\end{eqnarray*}
has good properties, because the final master equation
(\ref{Master equation of the amplification-feedback network}) is
of the traditional form of squeezing environment which has been
widely studied in quantum optics~\cite{Ficek1,Ficek2}.

{\it Derivation of Eq.~(\ref{Effective Hamiltonian of the coupled
TLR and input field system})}: Near the optimal point $n_g=1/2$,
the two energy levels of the charge qubit corresponding to $n=0,1$
are close to each other and far separated from higher-energy
levels. Thus, in this case, the charge qubit can be looked as a
two-level system~\cite{You,Makhlin,Clarke}. The Hamiltonian
$H_{qT}$ in Eq.~(\ref{Hamiltonian of the coupled qubit-TLR
system}) can be written as:
\begin{eqnarray*}
H_{qT}=-2E_C\left(1-2n_g\right)\tilde{\sigma}_z-E_J^0\cos\left(\pi\frac{\Phi_x}{\Phi_0}\right)\tilde{\sigma}_x+\omega_a
a^{\dagger}a,
\end{eqnarray*}
where the Pauli operators $\tilde{\sigma}_z$ and
$\tilde{\sigma}_x$ are defined by:
\begin{eqnarray*}
\tilde{\sigma}_z&=&|0\rangle\langle0|-|1\rangle\langle1|,\\
\tilde{\sigma}_x&=&|0\rangle\langle1|+|1\rangle\langle0|,
\end{eqnarray*}
and $|0\rangle$ and $|1\rangle$ are the charge states with the
Cooper pairs numbers $n=0,1$.

When $n_g=1/2$, the charge qubit is in the charge degenerate
point. In this case, $H_{qT}$ can be rewritten using the
eigenstates of the charge qubit as:
\begin{eqnarray*}
H_{qT}&=&\omega_a
a^{\dagger}a-E_J^0\cos\frac{\pi}{\Phi_0}\left[\Phi_e+\eta_T\left(a+a^{\dagger}\right)\right.\\
&&\left.+\eta_{\rm in}\left(b_{\rm in}+b_{\rm
in}^{\dagger}\right)\right]\sigma_z,
\end{eqnarray*}
where
\begin{eqnarray*}
\sigma_z=|+\rangle\langle+|-|-\rangle\langle-|,
\end{eqnarray*}
and
\begin{eqnarray*}
|+\rangle&=&\frac{1}{\sqrt{2}}|0\rangle+\frac{1}{\sqrt{2}}|1\rangle,\\
|-\rangle&=&-\frac{1}{\sqrt{2}}|0\rangle+\frac{1}{\sqrt{2}}|1\rangle.
\end{eqnarray*}
Since $\Phi_e=-\Phi_0/2$, we have
\begin{eqnarray*}
H_{qT}&=&-E_J^0\left[\sin\frac{\pi}{\Phi_0}\eta_T\left(a+a^{\dagger}\right)\cos\frac{\pi}{\Phi_0}\eta_{\rm
in}\left(b_{\rm in}+b_{\rm in}^{\dagger}\right)\right.\\
&&\left.+\cos\frac{\pi}{\Phi_0}\eta_T\left(a+a^{\dagger}\right)\sin\frac{\pi}{\Phi_0}\eta_{\rm
in}\left(b_{\rm in}+b_{\rm in}^{\dagger}\right)\right]\sigma_z\\
&&+\omega_a a^{\dagger}a.
\end{eqnarray*}
Expanding $H_{qT}$ to the first order of $b_{\rm in}+b_{\rm
in}^{\dagger}$ and the second order of
$\left(a+a^{\dagger}\right)$, we can rewrite $H_{qT}$ as:
\begin{eqnarray*}
H_{qT}&=&\omega_a
a^{\dagger}a-E_J^0\left[\frac{\pi\eta_T}{\Phi_0}\left(a+a^{\dagger}\right)\right.\\
&&\left.-\frac{\pi^3\eta_T^2\eta_{\rm
in}}{2\Phi_0^3}\left(a+a^{\dagger}\right)^2\left(b_{\rm in}+b_{\rm
in}^{\dagger}\right)\right]\sigma_z.
\end{eqnarray*}
Here, we have omitted the term
\begin{eqnarray*}
-\frac{\pi E_J^0 \eta_{\rm in}}{\Phi_0}\left(b_{\rm in}+b_{\rm
in}^{\dagger}\right)\sigma_z,
\end{eqnarray*}
which just leads to additional dephasing effects of the charge
qubit. Assume that the charge qubit always stays in the ground
state and omit the fast oscillating terms $a^{\dagger\,2}$, $a^2$
in the rotating wave approximation, we can obtain the effective
Hamiltonian of the TLR:
\begin{eqnarray*}
\tilde{H}_T&=&\omega_a
a^{\dagger}a-\frac{\pi^3E_J^0\eta_T^2\eta_{\rm
in}}{\Phi_0^3}a^{\dagger}a\left(b_{\rm in}+b_{\rm
in}^{\dagger}\right)\\
&&+\frac{\pi E_J^0\eta_T}{\Phi_0}\left(a+a^{\dagger}\right).
\end{eqnarray*}
The linear term
\begin{eqnarray*}
\frac{\pi E_J^0\eta_T\left(a+a^{\dagger}\right)}{\Phi_0}
\end{eqnarray*}
can be compensated by a classical driving field on the TLR, and
thus the effective Hamiltonian $H_T$ given in Eq.~(\ref{Effective
Hamiltonian of the coupled TLR and input field system}) can be
obtained.

\section*{Acknowledgment}
J. Zhang would like to thank Dr. H.~I. Nurdin for helpful
discussions and constructive suggestions for the revision of the
manuscript. The author would also like to thank Prof. K. Jacobs
and Dr. F. Sciarrino for providing the reference materials and
helpful comments.

\begin{biography}[{\includegraphics[width=1in,height=1.25in,clip,keepaspectratio]{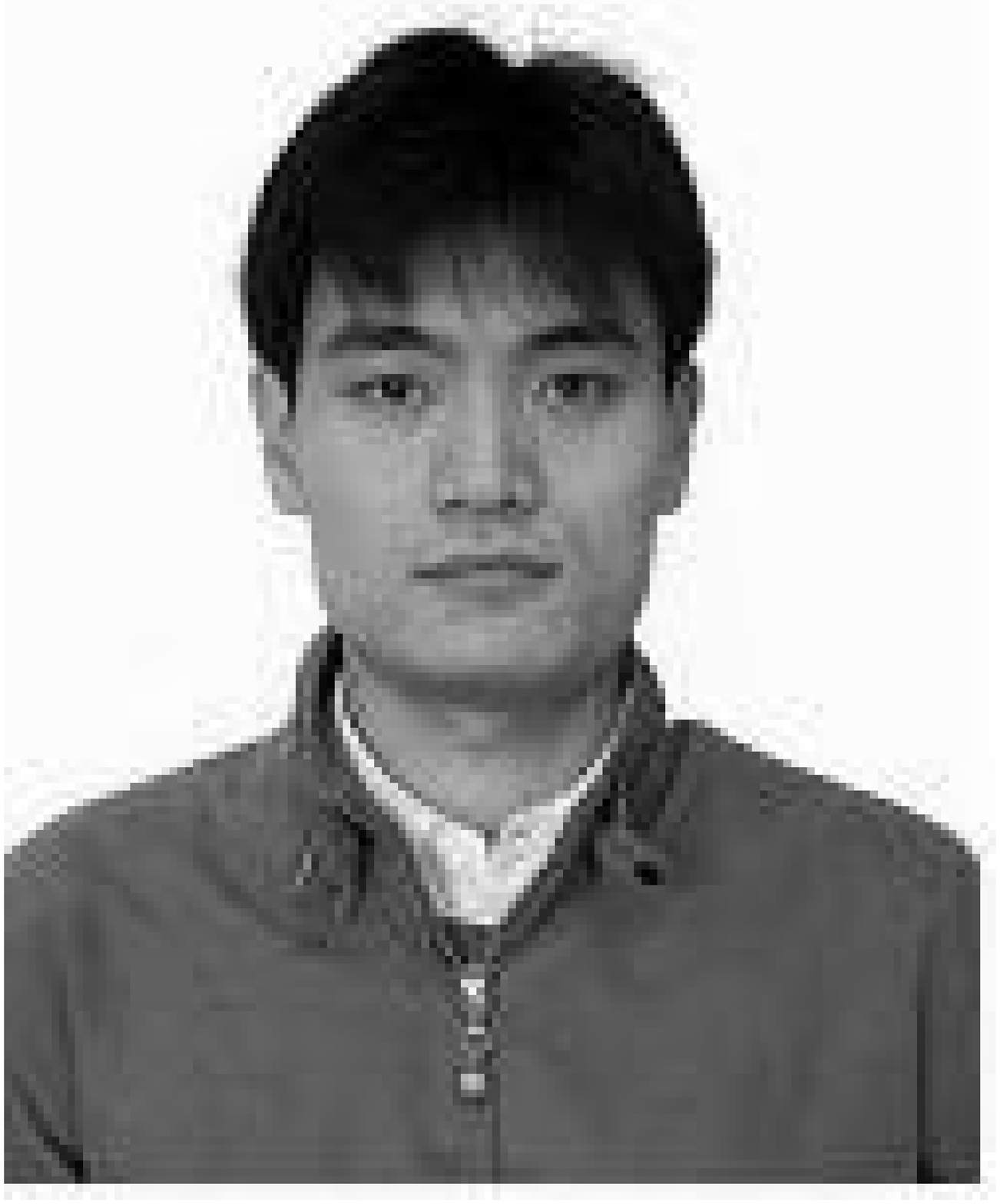}}]{Jing Zhang}
received his B.S. degree from Department of Mathematical Science
and Ph.D. degree from Department of Automation, Tsinghua
University, Beijing, China, in 2001 and 2006, respectively.

From 2006 to 2008, he was a Postdoctoral Fellow at the Department
of Computer Science and Technology, Tsinghua University, Beijing,
China, and a Visiting Researcher from 2008 to 2009 at the Advanced
Science Institute, the Institute of Physical and Chemical Research
(RIKEN), Japan. In 2010, he worked as a Visiting Assistant
Professor at Department of Physics and National Center for
Theoretical Sciences, National Cheng Kung University, Taiwan. He
is now an Associate Professor at the Department of Automation,
Tsinghua University, Beijing, China. His research interests
include quantum control and nano manipulation.
\end{biography}

\begin{biography}[{\includegraphics[width=1in,height=1.25in,clip,keepaspectratio]{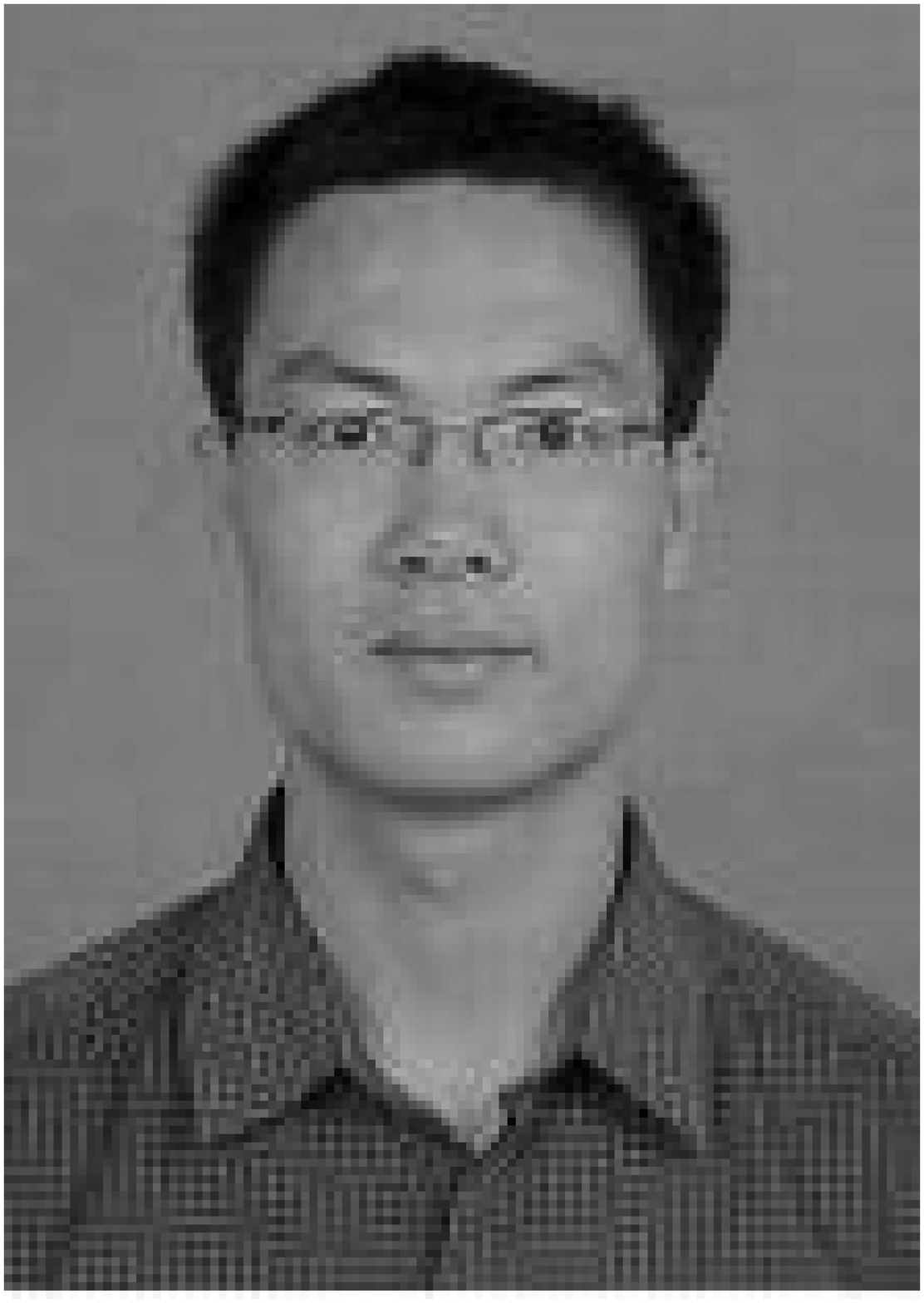}}]{Re-Bing Wu}
received his B.S. degree in Electrical Engineering and Ph.D.
degree in Control Science and Engineering from Tsinghua
University, Beijing, China, in 1998 and 2004, respectively.

From 2005 to 2008, he was a Research Associate Fellow at the
Department of Chemistry, Princeton University, USA. Since 2009, he
has been an Associate Professor at the Department of Automation,
Tsinghua University, Beijing, China. His research interests
include quantum mechanical control theory and nonlinear control
theory.
\end{biography}

\begin{biography}[{\includegraphics[width=1in,height=1.25in,clip,keepaspectratio]{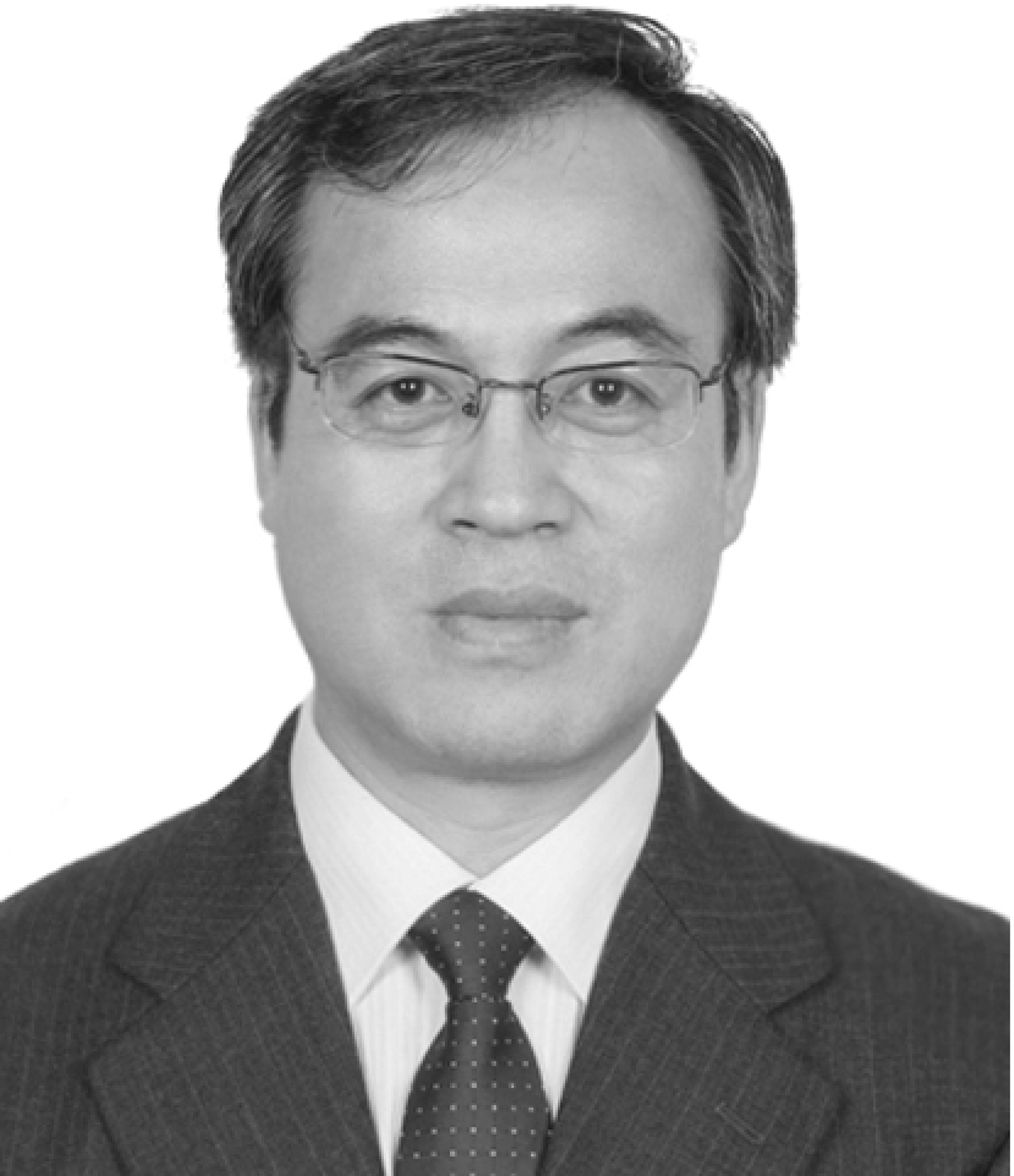}}]{Yu-xi Liu}
received his B.S. degree, Master degree and Ph.D. degree from
Department of Physics, Shanxi Normal University, Jilin University
and Beijing University in 1989, 1995 and 1998, respectively.

From 1998 to 2000, he was a Post-doctor at the Institute of
Theoretical Physics, the Chinese Academy of Sciences, China. From
2000 to 2002, he was a JSPS Postdoctoral fellow at the Graduate
University for Advanced Studies (SOKENDAI), Japan. From 2002 to
2009, he was a research scientist in the Institute of Physical and
Chemical Research (RIKEN), Japan.

Since 2009, he has been a Professor with Institute of
Microelectronics, Tsinghua University. His research interests
include solid state quantum devices, quantum information
processing, quantum optics and quantum control theory.
\end{biography}

\begin{biography}[{\includegraphics[width=1in,height=1.25in,clip,keepaspectratio]{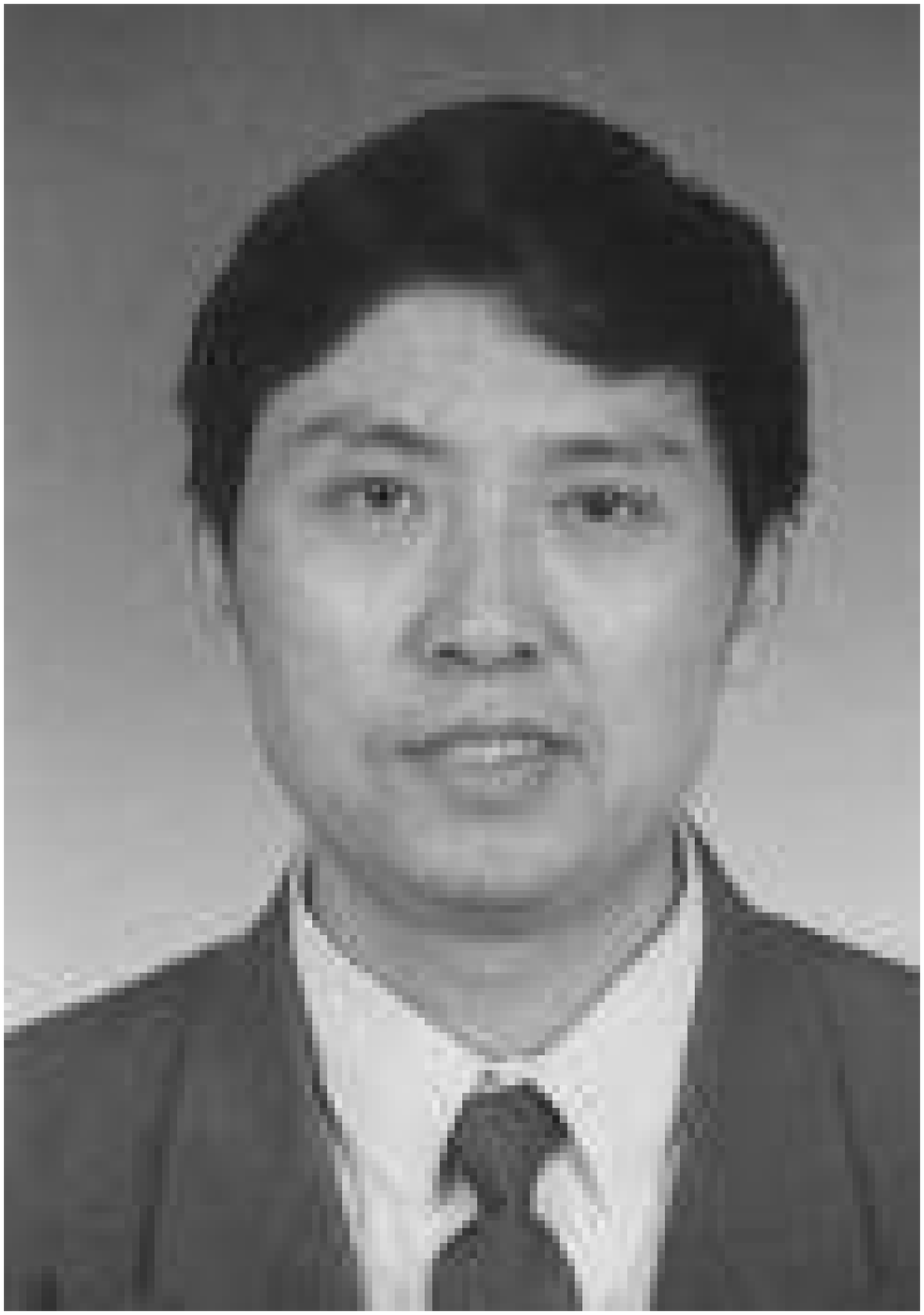}}]{Chun-Wen Li}
received his B.S. degree and Ph.D. degree from Department of
Automation, Tsinghua University in 1982 and 1989, respectively.

Since 1994, he has been a Professor with Department of Automation,
Tsinghua University. His research interests include nonlinear
control systems, inverse systems, CAD and simulation of nonlinear
systems, and robust control.

Prof. Li received the National Youth Prize in 1991 and the Prize
of Chinese Outstanding Ph.D. Degree Receiver in 1992.
\end{biography}

\begin{biography}[{\includegraphics[width=1in,height=1.25in,clip,keepaspectratio]{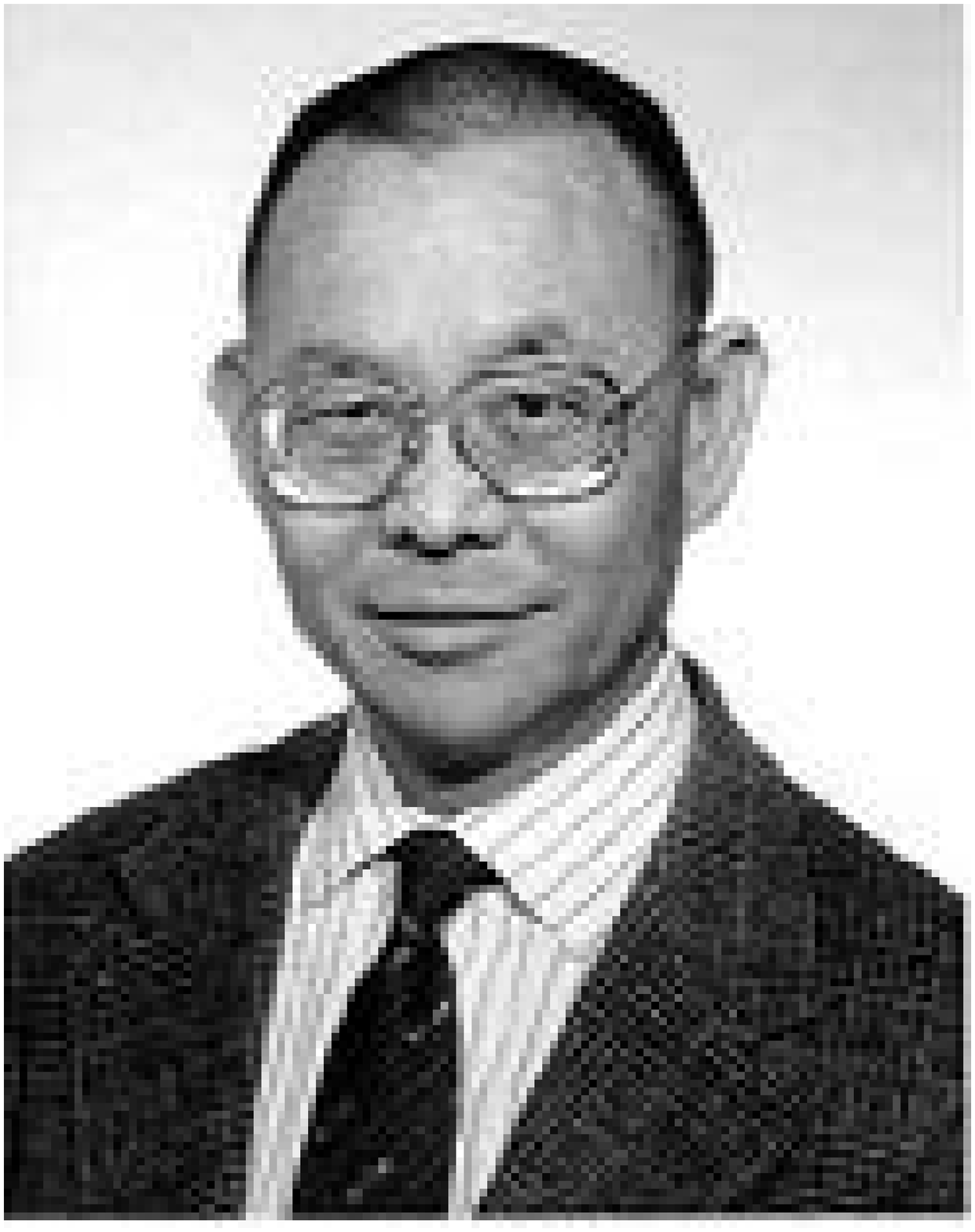}}]{Tzyh-Jong Tarn}
(M¡¯71-SM¡¯83-F¡¯85) received the D.Sc degree in control system
engineering from Washington University at St. Louis, Missouri, USA.

He is currently a Senior Professor in the Department of Electrical
and Systems Engineering at Washington University, St. Louis, USA.
He also is the director of the Center for Quantum Information
Science and Technology at Tsinghua University, Beijing, China.

An active member of the IEEE Robotics and Automation Society, Dr.
Tarn served as the President of the IEEE Robotics and Automation
Society, 1992-1993, the Director of the IEEE Division X (Systems
and Control), 1995-1996, and a member of the IEEE Board of
Directors, 1995-1996.

He is the first recipient of the Nakamura Prize (in recognition
and appreciation of his contribution to the advancement of the
technology on intelligent robots and systems over a decade) at the
10th Anniversary of IROS in Grenoble, France, 1997, the recipient
of the prestigious Joseph F. Engelberger Award of the Robotic
Industries Association in 1999 for contributing to the advancement
of the science of robotics, the Auto Soft Lifetime Achievement
Award in 2000 in recognition of his pioneering and outstanding
contributions to the fields of Robotics and Automation, the
Pioneer in Robotics and Automation Award in 2003 from the IEEE
Robotics and Automation Society for his technical contribution in
developing and implementing nonlinear feedback control concepts
for robotics and automation, and the George Saridis Leadership
Award from the IEEE Robotics and Automation Society in 2009. In
2010 he received the Einstein Chair Professorship Award from the
Chinese Academy of Sciences and the John R. Ragazzini Award from
the American Automatic Control Council. He was featured in the
Special Report on Engineering of the 1998 Best Graduate School
issue of US News and World Report and his research accomplishments
were reported in the ¡°Washington Times¡±, Washington D.C., the
¡°Financial Times¡±, London, ¡°Le Monde¡±, Paris, and the
¡°Chicago Sun-Times¡±, Chicago, etc. Dr. Tarn is an IFAC Fellow.

\end{biography}







\end{document}